\documentclass[referee,fleqn]{mnras}

\usepackage[T1]{fontenc}
\usepackage{ae,aecompl}

\usepackage{graphicx}	% Including figure files
\usepackage{amsmath}	% Advanced maths commands
\usepackage{amssymb}	% Extra maths symbols
\usepackage{pdflscape}
\usepackage{booktabs}
\newcommand{\angstrom}{\text{\normalfont\AA}}
\title[Effect of the isotropic collisions]{Effect of the isotropic collisions with neutral hydrogen  on the polarization of    the  CN solar molecule }

\author[]{S. Qutub$^{1}$, M. Derouich$^{1,2}$\thanks{derouichmoncef@gmail.com}, Y.N. Kalugina$^{3,4}$, H. Assiri$^{1}$, F. Lique$^{5}$ 
\\
$^{1}$Astronomy Dept, Faculty of Science, King Abdulaziz University, Jeddah, Saudi Arabia \\
$^{2}$ Sousse University,  ESSTHS, Lamine Abbassi street, 4011 H. Sousse, Tunisia \\
$^{3}$ 
Department of Optics and Spectroscopy, Tomsk State University, 36 Lenin av., Tomsk 634050, Russia \\
$^{4}$ Institute of Spectroscopy, Russian Academy of Sciences, Fizicheskaya St. 5, 108840 Troitsk, Moscow, Russia \\
$^{5}$ LOMC - UMR 6294, CNRS-Universit\'e du Havre, 25 rue P. Lebon, BP 1123, F-76063 Le Havre, France
}

\date{Accepted 2019 October 22. Received 2019 September 22; in original form 2019 July 04 }

\pubyear{2019}

\begin{document}
\label{firstpage}
 \pagerange{\pageref{firstpage}--\pageref{lastpage}}
\maketitle

\begin{abstract}
%context
%aims
 Our  work is concerned with the case of the solar molecule  CN which presents conspicuous profiles of scattering polarization.    We start by calculating accurate  PES for the singlet and triplet electronic ground states in order to characterize the collisions between    the CN molecule  in its $X \; ^2\Sigma$   state     and the hydrogen in its ground state $^2S$. The PES  are  included in the  Schr\"oodinger  equation to obtain the scattering matrix and the probabilities of collisions.    Depolarizing collisional rate coefficients are computed in the framework of the infinite order sudden approximation for  temperatures ranging from  $T= 2000$ K to $T= 15000$ K. Interpretation of the results and comparison between  singlet and triplet collisional rate coefficients are detailed.   We  show that, for typical photospheric hydrogen density ($n_{H}  =  10^{15}-10^{16}$    cm$^{-3}$), the  $X \; ^2\Sigma$   state  of CN is partially or completely depolarized by isotropic collisions.

%conclusions
\end{abstract}

%-----------------------------------------------
%-----------------------------------------------
\section{Introduction}

Observations of linear polarization close to the solar limb have revealed 
the existence, in the second solar spectrum (SSS), of prominent linear polarization signals due to molecular lines  (e.g. Mohan Rao  and  Rangarajan 1999; Gandorfer   2000; Faurobert  \& Arnaud 2003; Berdyugina \& Fluri  2004; Asensio Ramos \& Trujillo Bueno 2005; Mili\'c \& Faurobert 2012). In particular,   the CN  molecule, which   shows   weak lines in the usual (unpolarized) solar spectrum, presents conspicuous  peaks in the SSS  (Shapiro et al.  2011). The SSS is   the observational signature of the   polarization of the CN rotational levels which consists on population imbalances and quantum coherences among their    Zeeman sub-levels.

In  the solar photosphere, polarized CN levels  undergo  the effect  of  isotropic collisions between emitting or absorbing molecules   and nearby hydrogen atoms. 
Since collisions  are  isotropic, they tend to  partially or totally destroy the  polarization of CN  lines.    In addition, the Hanle effect of a solar magnetic field results in a partial  decrease of the      polarization of the CN states. Thus,  the depolarizing effects of the isotropic collisions and the Hanle  effect      are mixed in the same observable (the polarization state), which makes the interpretation    of the observed polarization  in terms of magnetic fields   complicated   because of the almost complete lack of collisional molecular  depolarization rates.    To derive  magnetic fields  from the interpretation of the  SSS, it is fundamental to firstly determine  the  depolarizing collisional rate   coefficients, and  then to include these data in the formalism of the formation of the polarized lines.

Over more than 40 years, vigorous efforts  were concerned with the calculation of the collisional ro-vibrational (de)excitation rates for interstellar molecules (e.g. Roueff  \& Lique 2013). The majority of the works were dedicated to the modelling of the molecular line profiles.  The effect of the collisional  excitation on the molecular polarization profiles is usualy overlooked. The literature in argument, there are no calculations of depolarization rates for solar molecular lines  by collisions with neutral hydrogen. 

Our intention in this work is to provide  new (de)polarization collisional rates for the CN molecule in its ground  state  $X \; ^2\Sigma$   which is very important in the solar polarization studies.   Computations of collisional rates occur in two steps. The first step is  the determination of potential energy surafces (PES) in order to characterize the interactions between the atom and the molecule. All the PES were obtained using the MOLPRO package  (e.g. Werner et al. 2010). 
 The second step is the study of the dynamics of collisions by solving the Schr\"oodinger  equation.   Dynamics calculations are   made possible thanks to the MOLSCAT  code (Hutson \& Green  1994).  
The depolarization cross-sections have been computed within the infinite-order-sudden (IOS) approximation
 for first 40 rotational levels and for kinetic energies ranging from 400 to  20000  cm$^{-1}$.
This allowed us to calculate depolarization rates of the    state  $X \; ^2\Sigma$    of the CN for temperatures between 2000 and 15000 K.

%in which     various types of collisions   are implemented and ready to be used. These types  involve  generally   structureless atoms like Helium as perturbers.  Howerver, one has to adapt the MOLSCAT code to be used for some more complex types of collisions with open-shell atoms  like the hydrogen in its ground state $^2S$.
%To discuss the accuracy of our results, we perform a full study of the  region of the  potential curve which  plays a decisive role in the evaluation of the depolarization rate.  

\section{Potential Energy Surfaces} \label{sec:PES}
\begin{figure}
 \centering{\includegraphics[width=5.5cm]{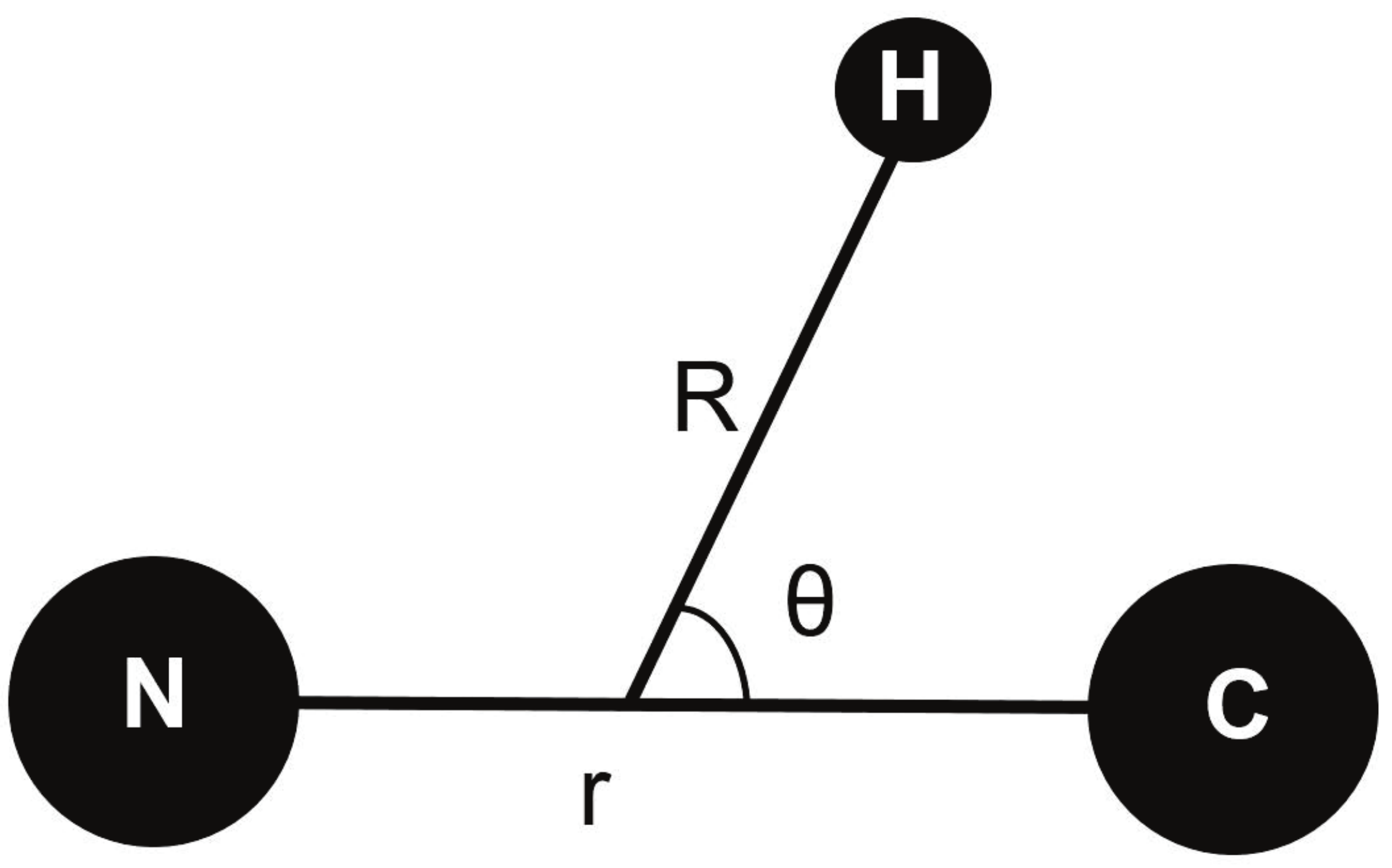}}
\caption{Coordinate system of the CN -- H interaction.}
\label{fig:coord}
\end{figure}

\noindent
In the present work, the coordinate system presented in
Fig.~\ref{fig:coord} was used. The center of coordinates coincides
with the center of mass of the CN molecule. Intermolecular vector $R$ connects the centre of mass of CN molecule and H atom. Angle $\theta$ defines the rotation of the hydrogen atom arount the CN molecule. Thus, the mutual orientation and position of the H atom is described by the intermolecular separation, $R$, and by angle $\theta$.

The CN molecule is assumed rigid with geometrical structure corresponding to the equilibrium: $r$~=~2.2144 ~$a_0$ (e.g. Yang et al. 2016). 

%%%%%%%%%%%%%%%%%%%%%%%%%%%%%%%%%%%%%%%%%%%%%
%%%%%%%%%%%%%%%%%%%%%%%%%%%%%%%%%%%%%%%%%%%%%

%%%%%%%%%%%%%%%%%%%%%%%%%%%%%%%%%%%%%%%%%%%%%
\begin{figure*}
\centering{\includegraphics[width=8.6cm]{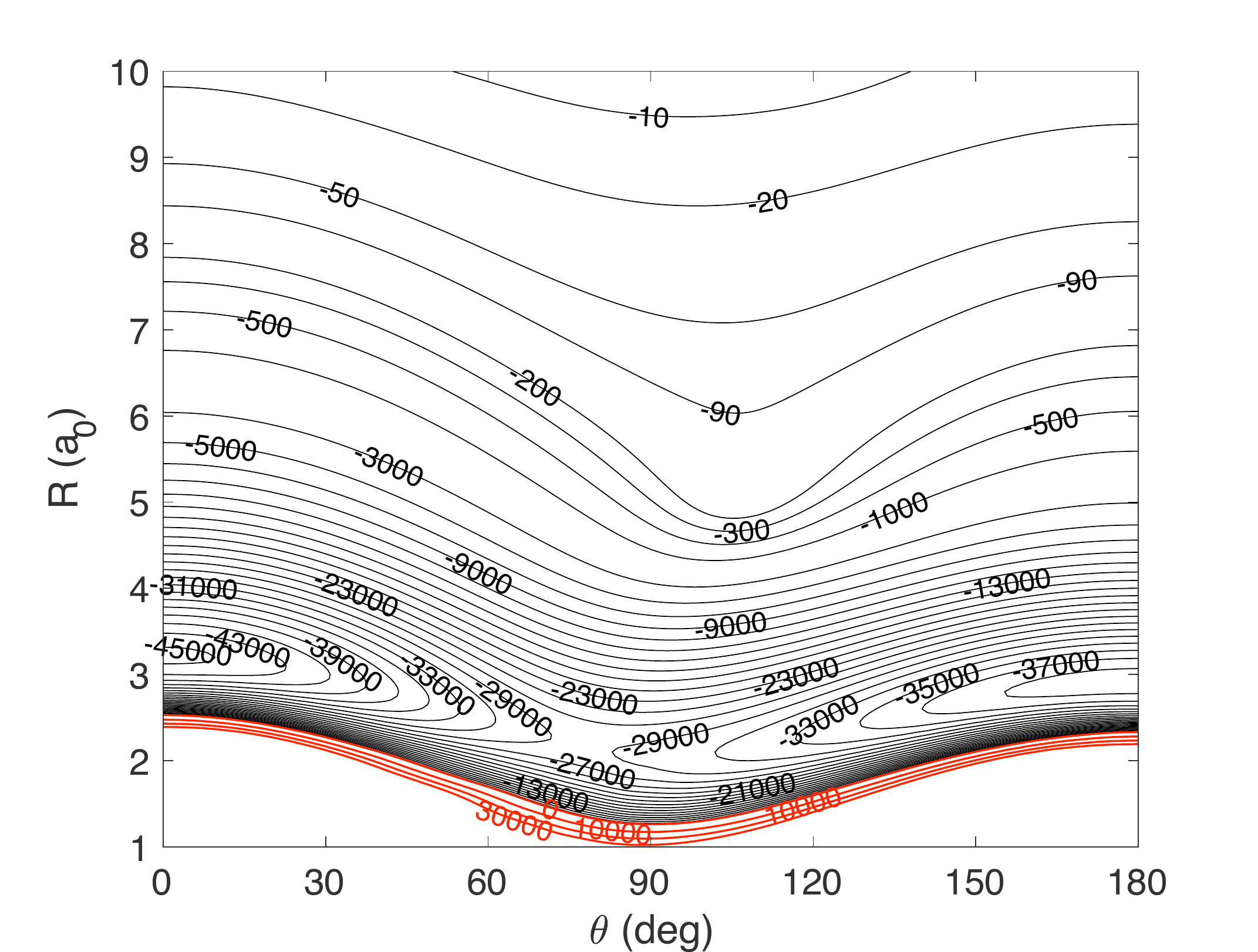}}
\hspace{-0.8cm}
\centering{\includegraphics[width=8.45cm]{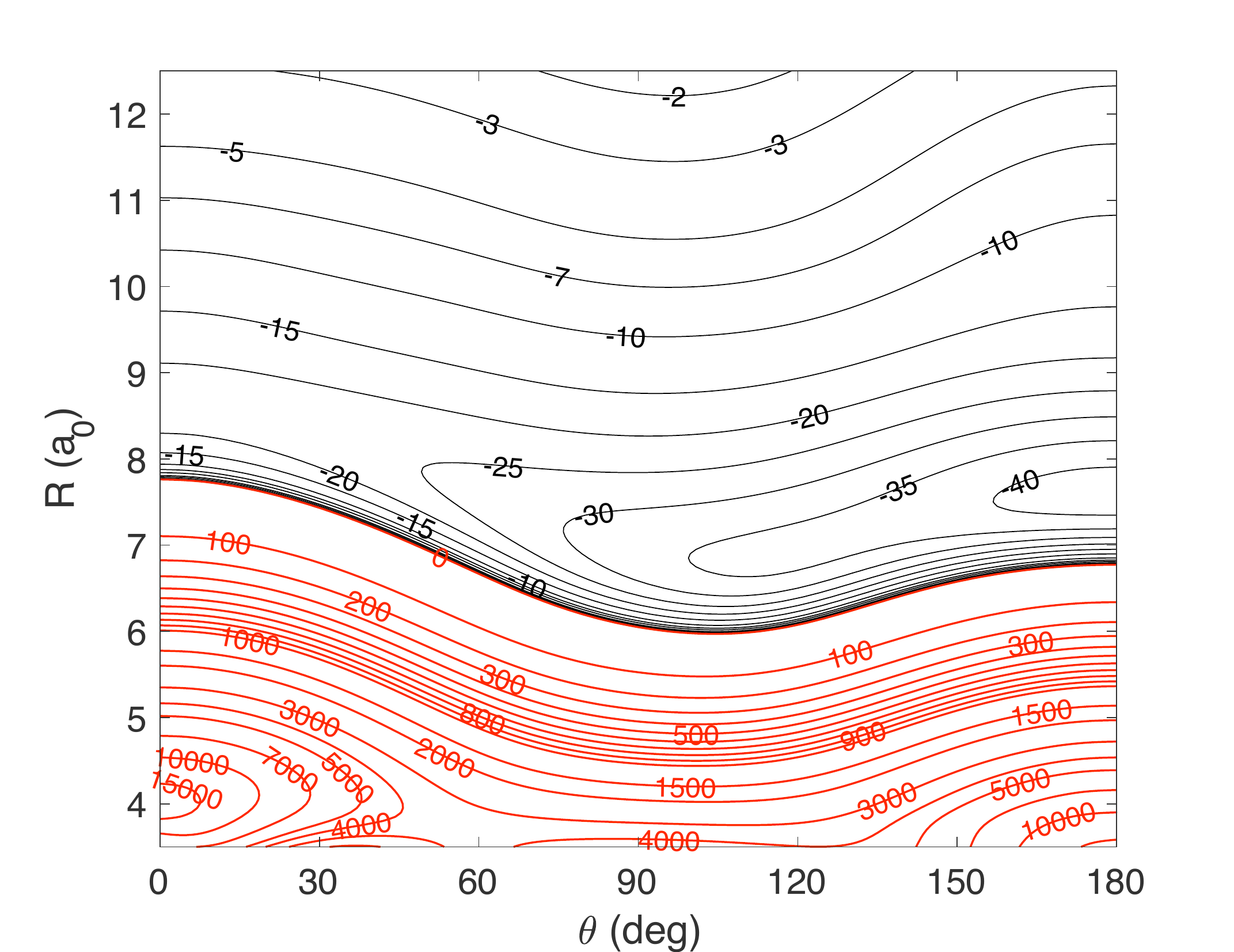}}
\vspace{-0.6cm}
\caption{2D potential energy surface for singlet state (left panel) and triplet state (right panel). Energy is in cm$^{-1}$.}
\label{fig:meth}
\end{figure*}
%%%%%%%%%%%%%%%%%%%%%%%%%%%%%%%%%%%%%%%%%%%%%

When CN molecule in the ground electronic state $^2\Sigma^+$ interacts with the hydrogen atom in $^2S$ ground electronic state the system can exist in two electronic states with total electronic spin $\vec{S_t}$ = $\vec{S}_{CN}$+$\vec{S}_H$. Thus, we have to obtain the potential energy surfaces for the singlet ($^1A$) and triplet ($^3A$) states.  

\emph{Ab initio} calculations of the PES for the singlet and triplet electronic states of CN -- H were carried out at multireference internally contracted configuration interaction (MRCI) (Wener 1988) level of theory. The size consistency was partially corrected using the Davidson (+Q) correction (Davidson 1977), the rest was corrected by subtracting the energy at $R$=100 $a_0$. The 1s core electrons of the Nitrogen and Carbon atoms were kept frozen. The active space consists of 10 electrons distributed in the 9 orbitals.   
The augmented correlation-consistent triple zeta (aVTZ) basis set (Dunning 1989) augmented by (3s, 2p, 1d) mid-bond functions (bf) (Williams et al. 1995) were used. The computations were performed using MOLPRO 2010 package (e.g. Werner et al. 2010).

For the singlet electronic state the intermolecular distance was varied from 1 to 60 $a_0$ giving 45 grid points. For the triplet electronic state the $R$ values were varied from 3.5 to 60 $a_0$ with total of 35 grid points. The angle $theta$ was varied from 0$^{\circ}$ to 180$^{\circ}$ with a step of 5$^{\circ}$. 
Due to high anysotropy of the $^1A$ potential the 2D-spline was employed for the representation of both PESs at any set ($R$, $\theta$).  The resulting potential energy surfaces for the singlet and triplet electronic states are presented in Fig.~\ref{fig:meth}.

There are two minima on the PES for the singlet state corresponding to the formation of HCN and HNC molecules. The HCN arrangement corresponds to the minimal structure with $\theta=0^{\circ}$ and $R=3.2$ $a_0$ and has the well depth $E$ = -45426 cm$^{-1}$. The HNC minimal structure corresponds to $\theta=180^{\circ}$ and $R=$2.92 $a_0$ and has a well depth $E$=-40098 cm$^{-1}$. The minimum for the triplet state occurs at $R$=7.6 $a_0$, $\theta=180^{\circ}$ and has an energy $E$=-42.40 cm$^{-1}$.

\begin{figure}
\centering
%\hspace{-0.7cm}
\includegraphics[width=10.5cm]{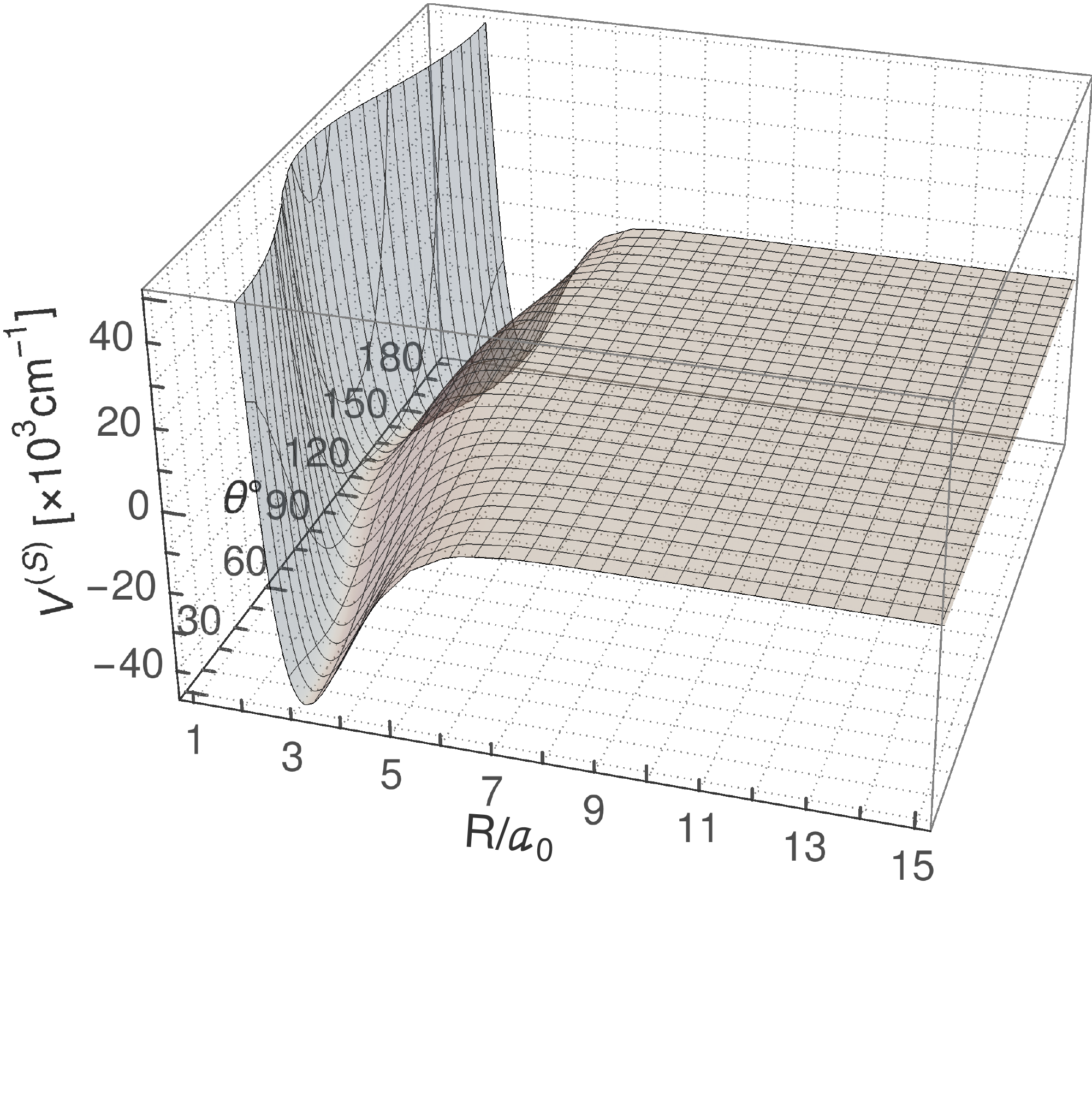} \\ \vspace{-2.0cm}
\includegraphics[width=10.0cm]{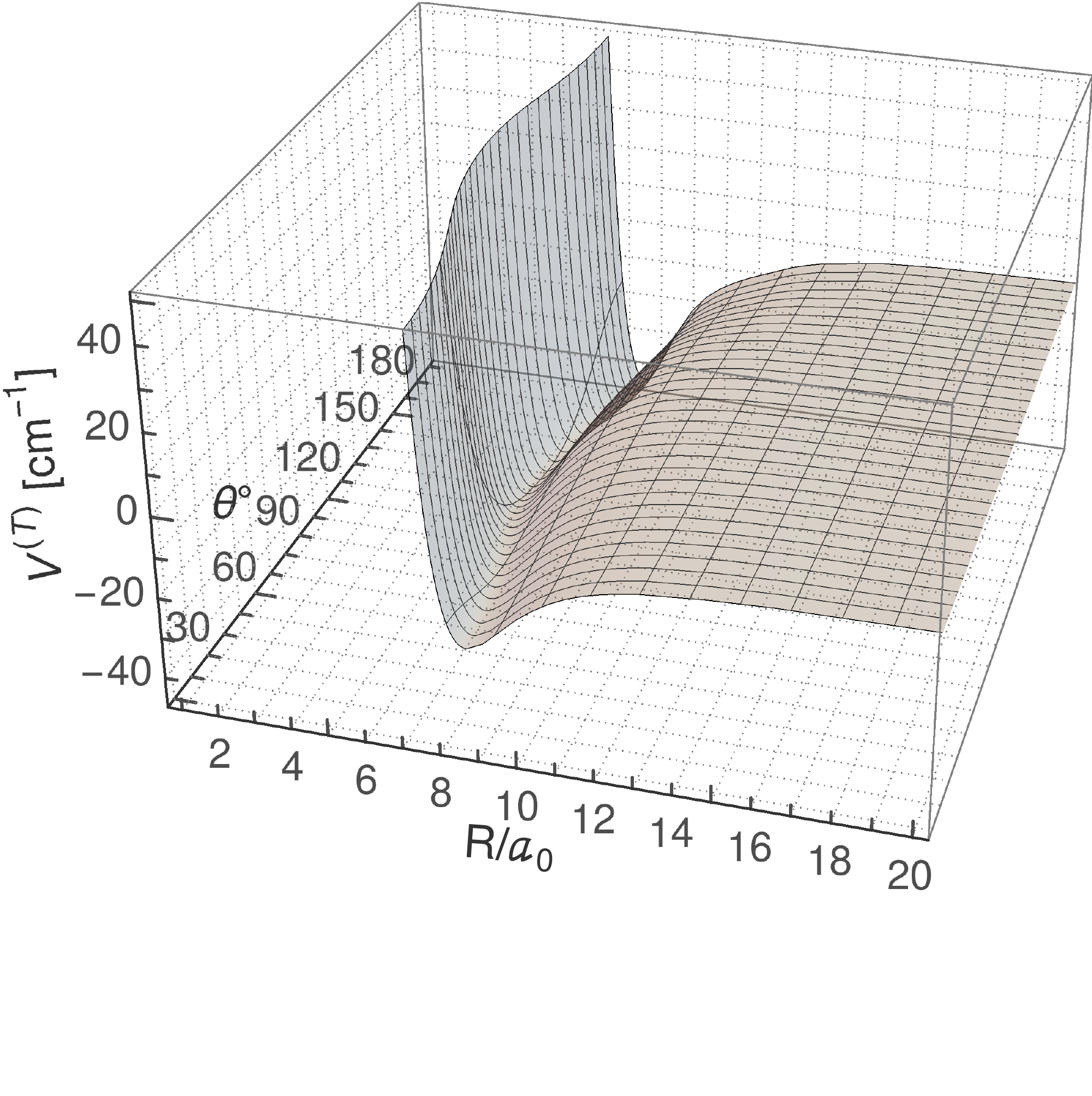} 
\vspace{-2.0cm}
\caption{3D plots of the singlet, $V^{(S)}$, and triplet, $V^{(T)}$, components of the CN-H potential, respectively.}
\label{fig:pot3d_s}
\end{figure}

In Figure \ref{fig:pot3d_s}, we respectively show 3-dimensional plots of singlet and triplet components of the potential energy for the CN-H system.

%%%%%%%%%%%%%%%%%%%%%%%%%%%%%%%%%%%%%%%%%%%%%
%%%%%%%%%%%%%%%%%%%%%%%%%%%%%%%%%%%%%%%%%%%%

\section{Theoretical background}
\subsection{Basis of irreducible tensorial operators}
Physical interpretation of the solar polarization  requires suitable description of the internal states of the emitting/absorbing molecules  with  the density matrix formalism expressed on the basis of irreducible tensorial operators    (e.g. Sahal-Br\'echot 1977; Trujillo Bueno 2002;  Landi Degl'Innocenti  \& Landolfi 2004). The CN  states are described by the density matrix elements $ \rho_{q}^{k} (j) $ where $j$ is the  molecular angular momentum, $k$ is the tensorial order ($0  \le  k  \le 2 j$)
 and   $q$ quantifies the coherence between the sublevels ($-k  \le  q  \le k$).  The circular polarization of the molecule   is quantified  by the density matrix  elements  with odd  $k$ ($^{j}\rho_{q}^{k=1} $, $^{j}\rho_{q}^{k=3} $, etc.), while the linear polarization is quantified with the even ones: 
   $^{j}\rho_{q}^{k=2} $, $^{j}\rho_{q}^{k=4} $, etc. The  intensity of the transition  involving the $j$-level is given by the density matrix element  $^{j}\rho_{q}^{k=0} $. In order to study the SSS, one has to calculate the   density matrix  elements  with even  tensorial orders $k$=0, 2, 4, etc. In addition, it is to be noticed that $^{j}\rho_{q}^{k=0} $ is proportional to the $j$-level population. 
   
 In   studies concerned with the analysis   of only the intensity spectrum of the light,     only the element  $^{j}\rho_{q}^{k=0} $ is the unknown to be determined and one needs solely the collisional rates with  $k=0$.  These rates are similar to the usual collisional excitation rates which are typically calculated and included in the radiative transfer codes synthesising the ordinary intensity profiles. Rates with  $k \ge 0$ are needed for the cases of spectropolarimetric studies, i.e. where the goal is to synthesise  not only the intensity profiles  but also the polarization profiles. One needs to adopt   a formalism treating the interaction between an open-shell systems which takes in consideration the cases where $  k > 0$   corresponding to the effect of the collisions on the polarized light.
Note that the isotropy of the velocity distribution of the hydrogen atoms implies that the  polarization transfer rates are $q$-independent  (e.g. Sahal-Br\'echot 1977; Derouich et al. 2003; Derouich et al. 2007).

\subsection{Coupling scheme approach}
We are interested  in the electronic $^2\Sigma^+$ state of   the CN solar molecule.  
The CN   levels can be described in the Hund's  case (b) limit. 
The fine structure levels are labeled by $N$ and $j$, where $N$ is
the rotational angular momentum and $j$ the total molecular
angular momentum given by $j=N+S_d$ where   $S_d=1/2$ represents the spin of CN in its $^2\Sigma^+$ state. Thus,  $j=N \pm 1/2$.
CN molecule in the $^2\Sigma^+$ state collides with   hydrogen atom   in its ground state $^2S$. The spin of the hydrogen is $S_a=1/2$, thus the collision results in    producing a singlet state $^1A'$ with total spin $S_t=0$ and a triplet state $^3A'$ with $S_t=1$.

Corey \& Alexander (1985)  and  Corey et al. (1986) have studied the general case concerned with  cross-sections for collisions between open-shell systems; however they did not take into account   the effects of the collisions on the molecular polarization.   Corey \& Alexander (1985)  and  Corey et al. (1986)  found that in the expression of each cross-section  contains the effect of   different comopnents of the interaction potentials. For instance,   the   cross-section associated to a singlet state includes not only the singlet comopnent  of the interaction potential but also the triplet component, in addition to   an interference term arising from the open-shell nature of both colliding systems.  Corey \& Alexander (1985)  and  Corey et al. (1986) showed that, in the   infinite-order-sudden (IOS) approximation, the excitation  cross-sections can   be written as a linear combination of   IOS   cross-sections (see also Goldflam et al. 1977).  Interestingly, in the expression of the IOS cross-section one has also to take into account the contribution of  interaction potentials with different spin values.

We follow  the formalism presented in  Corey \& Alexander (1985)  and  Corey et al. (1986), and we apply it to obtain the expression of the polarization transfer and depolarization cross-sections $\sigma^k(Nj \to N'j')$  due to the isotropic collisions between the CN  molecule  with an open-shell perturber like the hydrogen. We adopt  the  IOS approximation which can be well justified especially for sufficiently high temperatures (see e.g. Lique et al. 2007).  In fact, as we are interested in the solar context, where the temperatures and the kinetic energies of collisions are high, one can expect that some simplification regarding the coupling effects should be invoked in order to obtain results with  acceptable accuracy in resonable computing time. Let us notice also that this is the first work which is intentend to determine depolarization rates by collisions between a molecule in an open shell state and the hydrogen atom in its  open shell ground state.   Our approximate approach can be summarized by the following indications:
\begin{itemize}
\item We start by precisely calculating the interaction potentials associated to the singlet   $^1A'$ with total spin $S_t=0$ and the triplet state $^3A'$ with $S_t=1$. 
\item Then, the Schr\"odinger equation describing the dynamics of collisions is solved for each  potential to obtain the corresponding  singlet and triplet IOS cross-sections $\sigma(0 \to L)$.  
 \item  In these conditions, for each value of the total spin,  we assume that the depolarization cross-sections for a tensorial order $k$ are given by the same expression established previousely in the case of an openshell molecule with a spinless atom. This assumption allows us   to  factorize the $\sigma^k(Nj \to N'j')$ into a product of terms
involving the geometrical factors and coupling scheme effects   and a term  given as a linear combination of  the IOS cross-sections $\sigma(0 \to L)$ (e.g. Corey \& Smith 1985): 
\begin{eqnarray}\label{eq_5}
\sigma^k(Nj \to N'j') = \sum_{\scriptstyle L} 
(-1)^{k+L+j+j'+1}  
 \left\{ \begin{array}{ccc} 
 j  & j' & L   \\
 j'  & j &k
\end{array}
\right\}    \nonumber \\
(2N+1)(2N'+1)(2j'+1)(2j+1)  \times \left\{ \begin{array}{ccc} 
 N  & N' & L  \\
j' & j &S_d
\end{array}
\right\}^2   \nonumber \\
 \left( \begin{array}{ccc} 
  N'& N  & L \\
 0 &  0 &  0 
\end{array}
\right)^2  \sigma(0  \to L) \,.
\end{eqnarray}
\end{itemize}
The depolarization cross-sections are defined by (e.g. Derouich et al. 2003; Landi Degl'Innocenti \& Landolfi 2004, Dagdigian \& Alexander 2009):
\begin{eqnarray}\label{eq_6}
\sigma^k(Nj) =\sigma^0(Nj \to Nj)- \sigma^k(Nj \to Nj) \,.
\end{eqnarray}
In order to obtain the total cross-section averaged over the spin one has:
\begin{eqnarray}\label{eq_7}
\sigma^k(Nj \to N'j') &=&  \frac{1}{4} \times [3 \times \sigma^k(Nj \to N'j'; \; ^3A')+\sigma^k(Nj \to N'j'; \;  ^1A')] \,.
\end{eqnarray}

The  depolarization rates, $ D^k(Nj, T)$,  of the  level $(Nj)$  due to elastic collisions 
 and   the polarization transfer rates, $ D^k(Nj \to v'N'j', T)$,  between the levels  $Nj$ and $v'N'j'$ are given by   integration over Maxwellian distribution of relative kinetic energies (or relative velocities). In addition,
\begin{eqnarray} \label{gammak}
 D^k(Nj) =D^0(Nj \to Nj)- D^k(Nj \to Nj)
\end{eqnarray}
which means that $D^0(Nj) =0$ by definition.  
 
 If  $S_d=S_t$=0, thus $j=N$, $j'=N'$,  and $j+j'=N+N'$ is even, one can demonstrate that (e.g. Derouich 2006; Lique et al. 2007):
\begin{eqnarray}\label{eq_8}
\sigma^k(Nj \to N'j') = \sigma^k(j \to j')=\sum_{\scriptstyle L>0}^{j+j'} 
(-1)^{k}  (2j'+1) (2j+1)  
 \left\{ \begin{array}{ccc} 
 j' & j' & k   \\
 j  & j &L
\end{array}
\right\}  \nonumber \\
 \left( \begin{array}{ccc} 
  j& L  & j' \\
 0 &  0 &  0 
\end{array}
\right)^{2} \sigma(v0 \to vL)
\end{eqnarray}
and in the case where  $k=0$, one finds (e.g. Derouich 2006):
\begin{eqnarray}\label{eq_9}
\sigma^0(Nj \to N'j') =
  \sqrt{\frac{2j+1}{2j'+1}} \sigma(Nj \to N'j')   \,.
\end{eqnarray}

 \subsection{Statistical equilibrium equations}
Physical interpretation of the observed polarization  requires the solution of the coupling between   polarized radiative transfer in the solar atmosphere and  statistical equilibrium equations.    In such situation, description of the internal states of the emitting/absorbing molecule   in the density matrix formalism expressed on the basis of irreducible tensorial operators  is shown as a most suitable  (e.g. Sahal-Br\'echot 1977; Landi Degl'Innocenti  \& Landolfi 2004). 
The contribution of the  isotropic collisions to the statistical equilibrium equations is:      
\begin{eqnarray} \label{eq_ch3_17}
\big(\frac{d \; ^{j}\rho_q^{k}}{dt})_{coll} & = & - D^k(j, T) \; ^{j}\rho_q^k \nonumber \\
&& - ^{j}\rho_q^k \sum_{j' \ne j}  \sqrt{\frac{2j'+1}{2j+1}} D^0 (j \to  j', T) \\
&& + \sum_{j' \ne j}  
D^k(j' \to  j, T) \;  ^{vj'}\rho_q^k  \nonumber 
\end{eqnarray}
The quantities to be computed  are the  density matrix elements 
$^{j}\rho_q^k$.    

  $ D^k(j, T)$ and $D^k(j' \to  j, T)$ should be  calculated independently    to enter  the statistical equilibrium equations.  It is to be noticed that, in the solar physical conditions, the   polarization transfer rates   for vibrational relaxation are  smaller than these for pure rotational relaxation by about three orders of magnitude. Therefore, it is convenient to neglect the effect of the transfer of polarization between  different vibrational states.   

\section{Results and discussions}

We separately feed the singlet and triplet parts of the potential  into MOLSCAT code which determines the dynamics of the colliding system by solving the corresponding Schr\"odinger equation and returns the scattring matrix and cross-sections. To obtain the depolarization and transfer of polarization rates, we thermally average the resultant cross-sections over the kinetic energy distributions of the colliding partners for temperatures ranging from $2000$ K to $15000$ K using the relations (see e.g.~Flower 2003),
\begin{eqnarray} \label{eq:thermal_xsection}
\langle \sigma^k v \rangle = \left( \frac{8}{\pi \mu k_B^3 T^3} \right)^{1/2} \!\! \int_0^\infty \!\! \sigma^k (E) \exp \left(-\frac{E}{k_B T}\right) E dE,
\end{eqnarray}
 where  $E$ is the kinetic energy of the incident atom with respect to the upper level and $\mu$ is the reduced mass of the colliding system.
For this purpose, we consider collision energies ranging from  400 \, cm$^{-1}$ to  20000 \, cm$^{-1}$. For energies larger than  20000 \, cm$^{-1}$, we extrapolate the cross-sections as their variation with   energy becomes  almost linear for sufficently large energies.
We then calculate the singlet and triplet contibutions to the transfer of polarization and depolarization rates using Eqs.~\ref{eq_5}, \ref{eq_6} and \ref{gammak}.
The results of our calculations are shown in Sections~\ref{sec:pol_transf_rate} and \ref{sec:depol_rate} below.

\subsection{Transfer of polarization cross-sections and rates } \label{sec:pol_transf_rate}

\begin{figure}
\centering
%\hspace{-0.7cm}
\includegraphics[width=8.0cm]{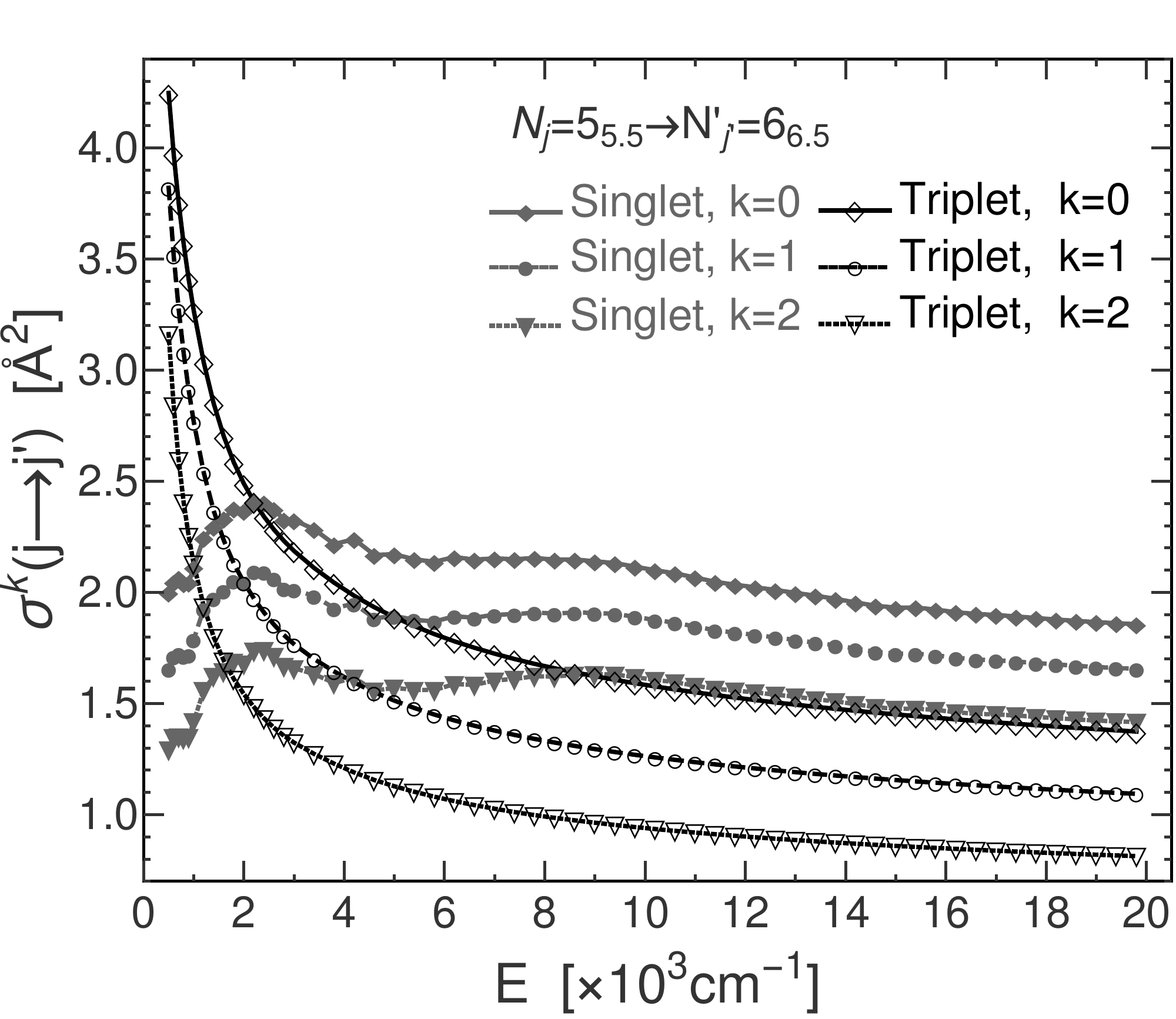}  
\includegraphics[width=8.0cm]{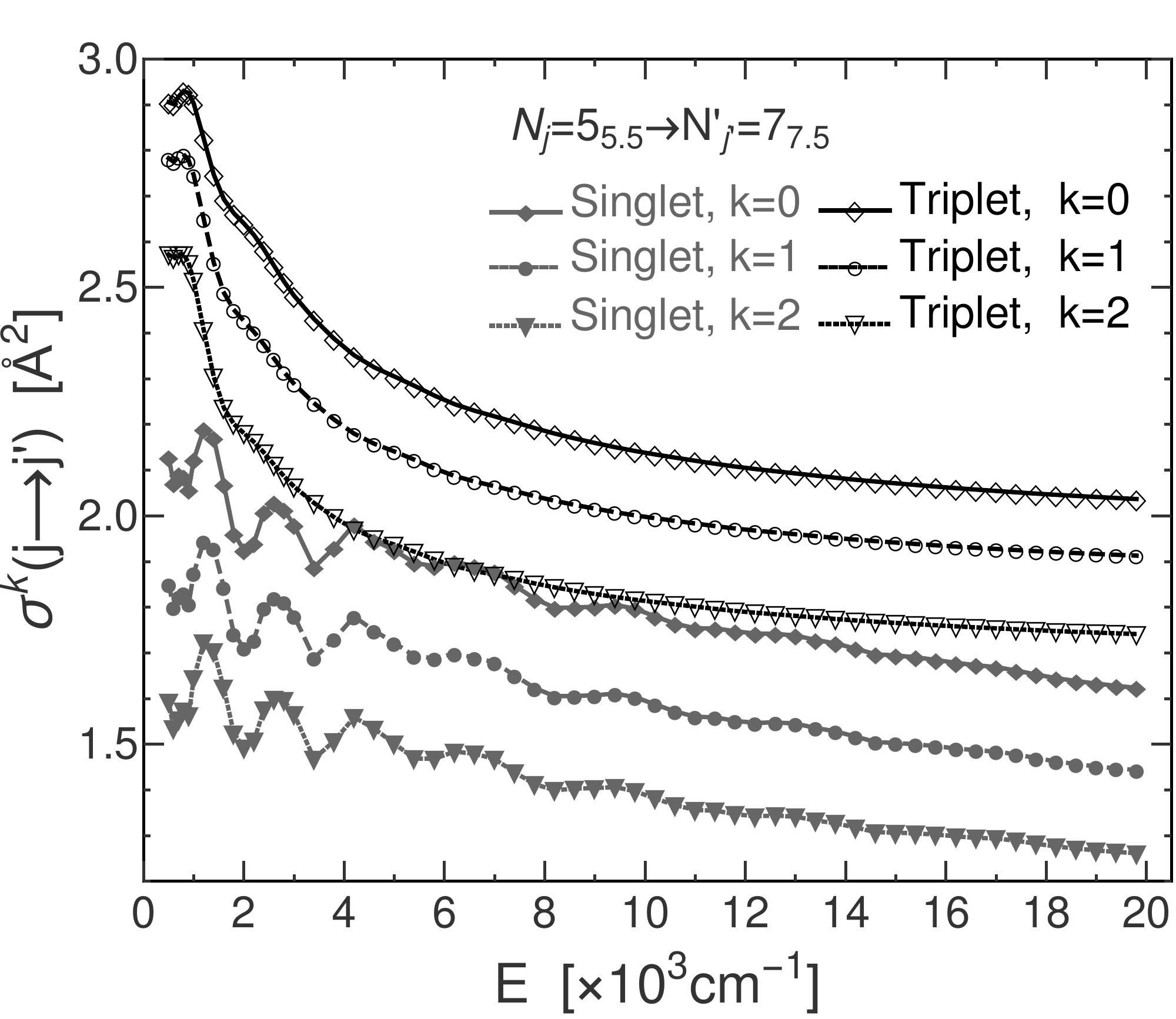}  \\
\includegraphics[width=8.0cm]{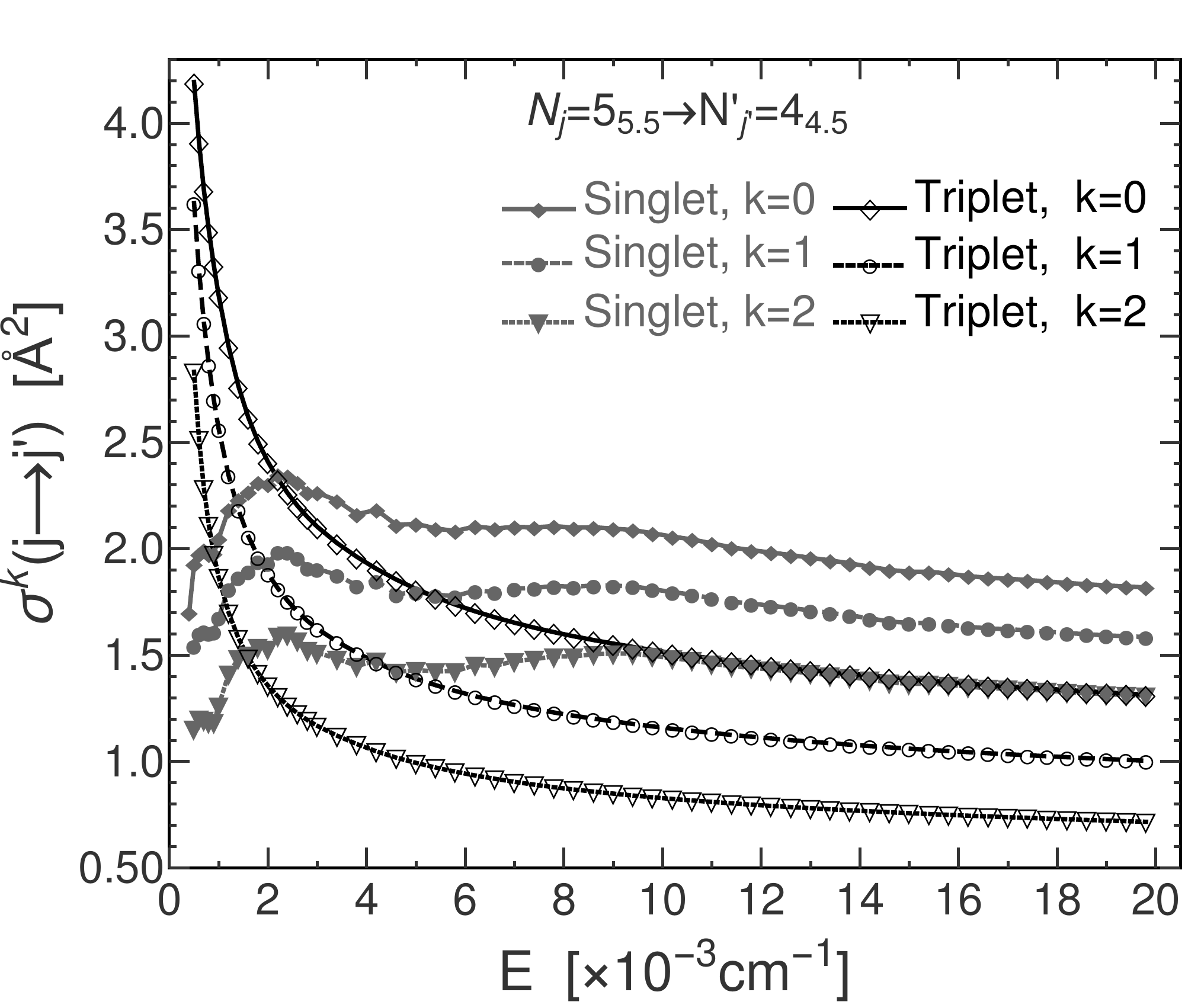}   
\includegraphics[width=8.0cm]{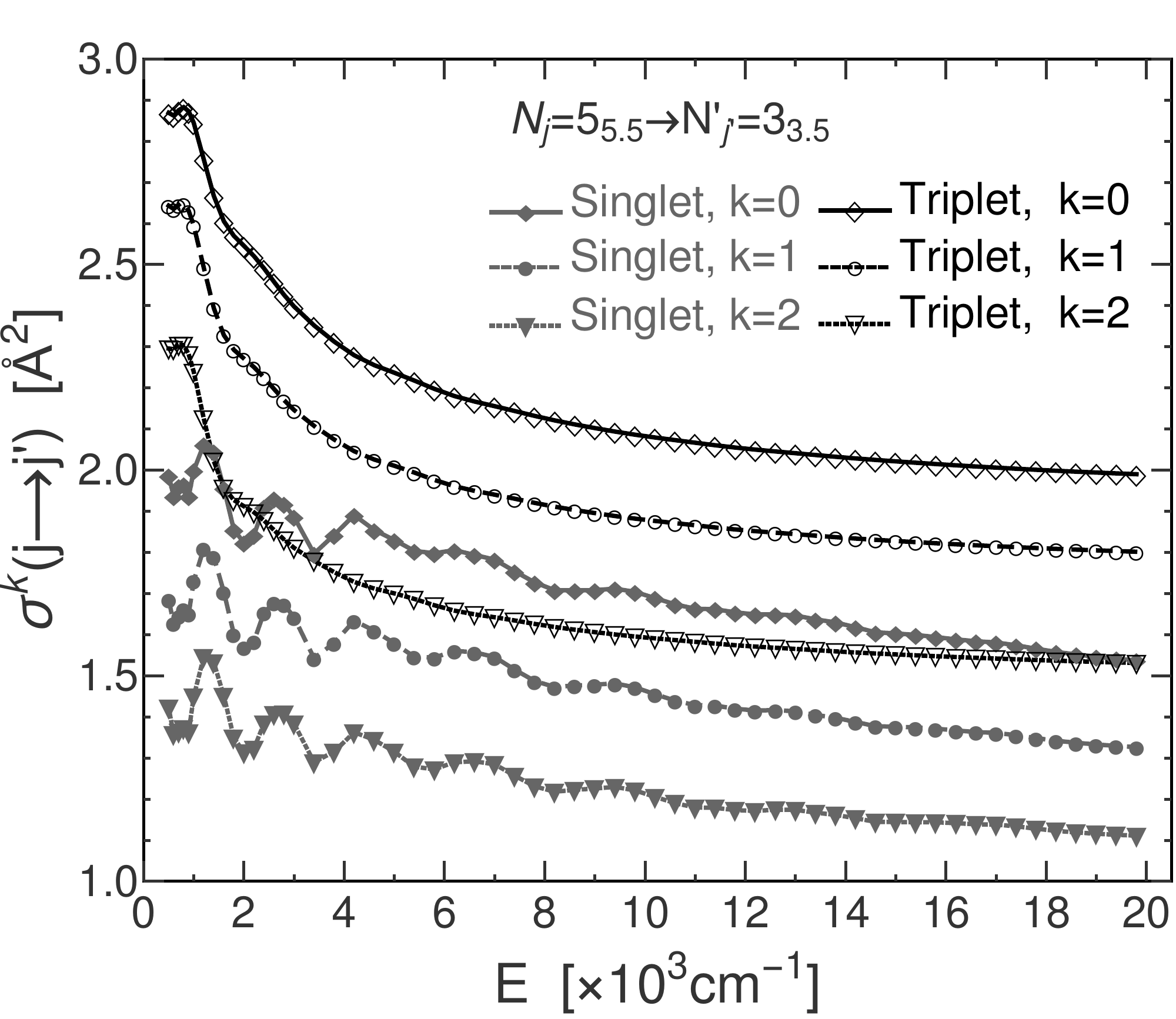}
\caption{Variation of the upward (upper panels) and downward (lower panels) transfer of polarization cross-sections with energy for the rotational level $N_j=5_{5.5}$. The singlet and triplet contributions are respectively shown by the gray and black curves for the population, $k=0$ (diamonds), orientation, $k=1$ (circles), and alignment, $k=2$ (down-triangles).}
 \label{fig:sigma_deltajm1}
\end{figure}

In  Figure~\ref{fig:sigma_deltajm1}, we show the energy variation of the upward (upper panels)
%[\ref{fig:sigma_deltajm1} and \ref{fig:sigma_deltajm2}] 
and downward (lower panels)
%[\ref{fig:sigma_deltaj1} and \ref{fig:sigma_deltaj2}]
 transfer of polarization cross-sections for the rotational level $N_j=5_{5.5}$. The contributions of the singlet and triplet parts of the potential are represented by the gray and black curves, respectively. 
%
%
%%%%%%%%%%%%%%%%%%%%%%%%%%%%%%%%%%%%%%%%%%%%
%%%%%%%%%%%%%%%%%%%%%%%%%%%%%%%%%%%%%%%%%%%%
%
%
We remark that the cross-sections decrease as $k$ increases.
Further, 
the upward (excitation) and the downward (de-excitation) transfer cross-sections with the same $\vert \Delta N \vert =\vert N' - N \vert$ from a given level have a very similar behavior; however the upward transfer cross-sections are slightly larger than the downward ones. The difference between the two increases with increasing $\vert \Delta N \vert$.
Furthermore, we notice that the size of the singlet and triplet contributions to the transfer of polarization cross-sections tends to alternate as $\vert \Delta N \vert$ increases.

\begin{figure}
\centering
%\hspace{-0.7cm}
\includegraphics[width=8.0cm]{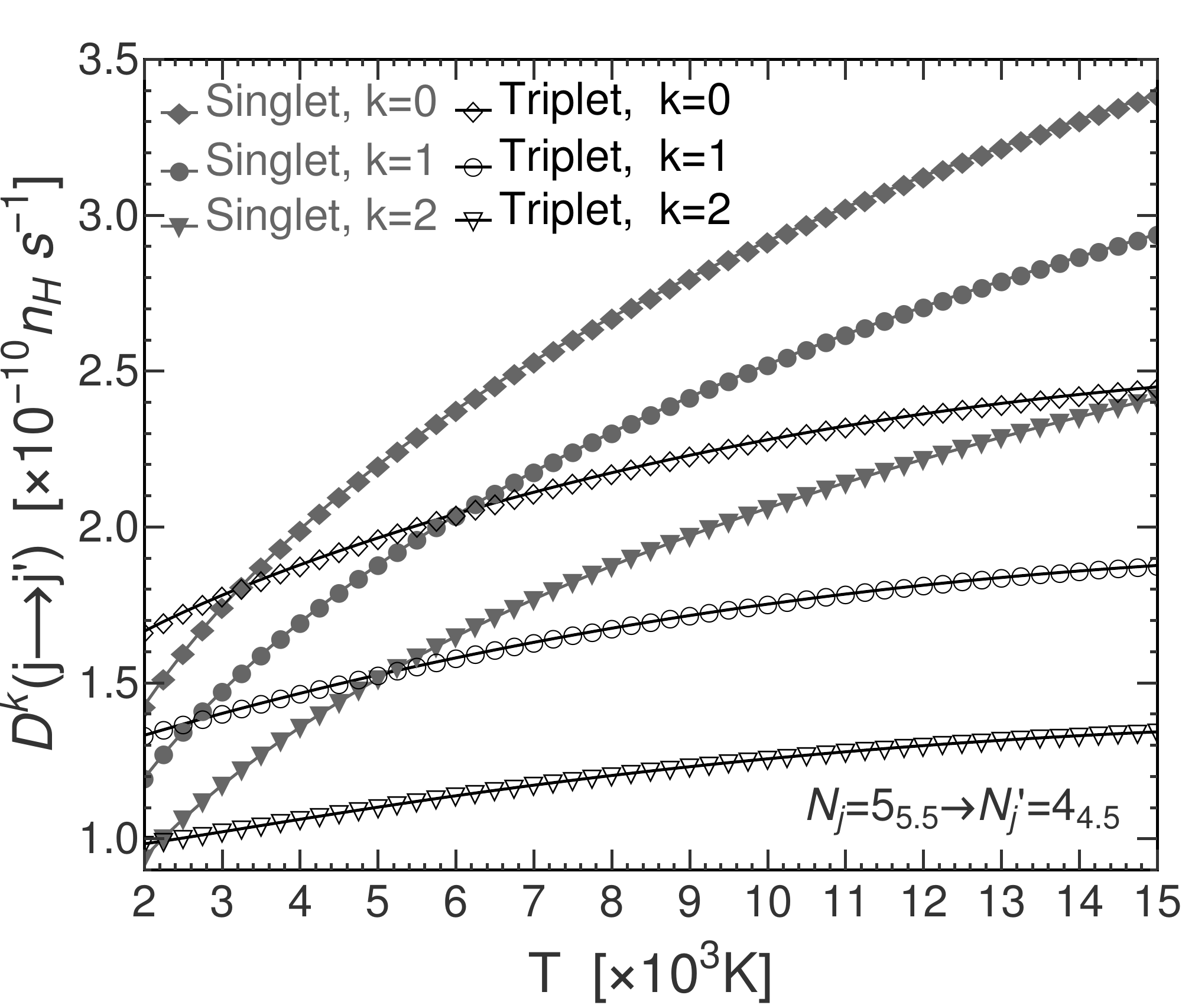} 
\includegraphics[width=8.0cm]{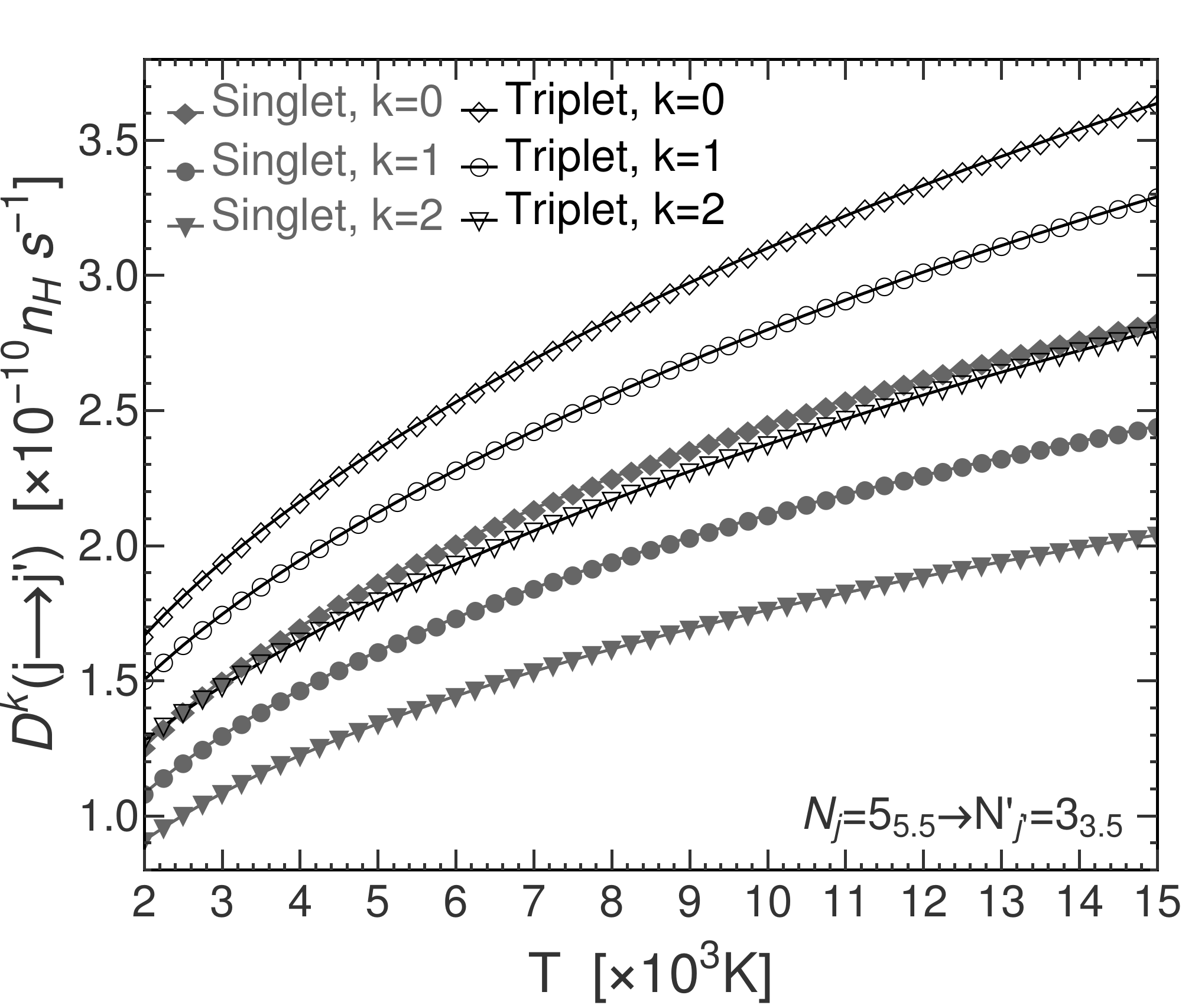}\\
\includegraphics[width=8.0cm]{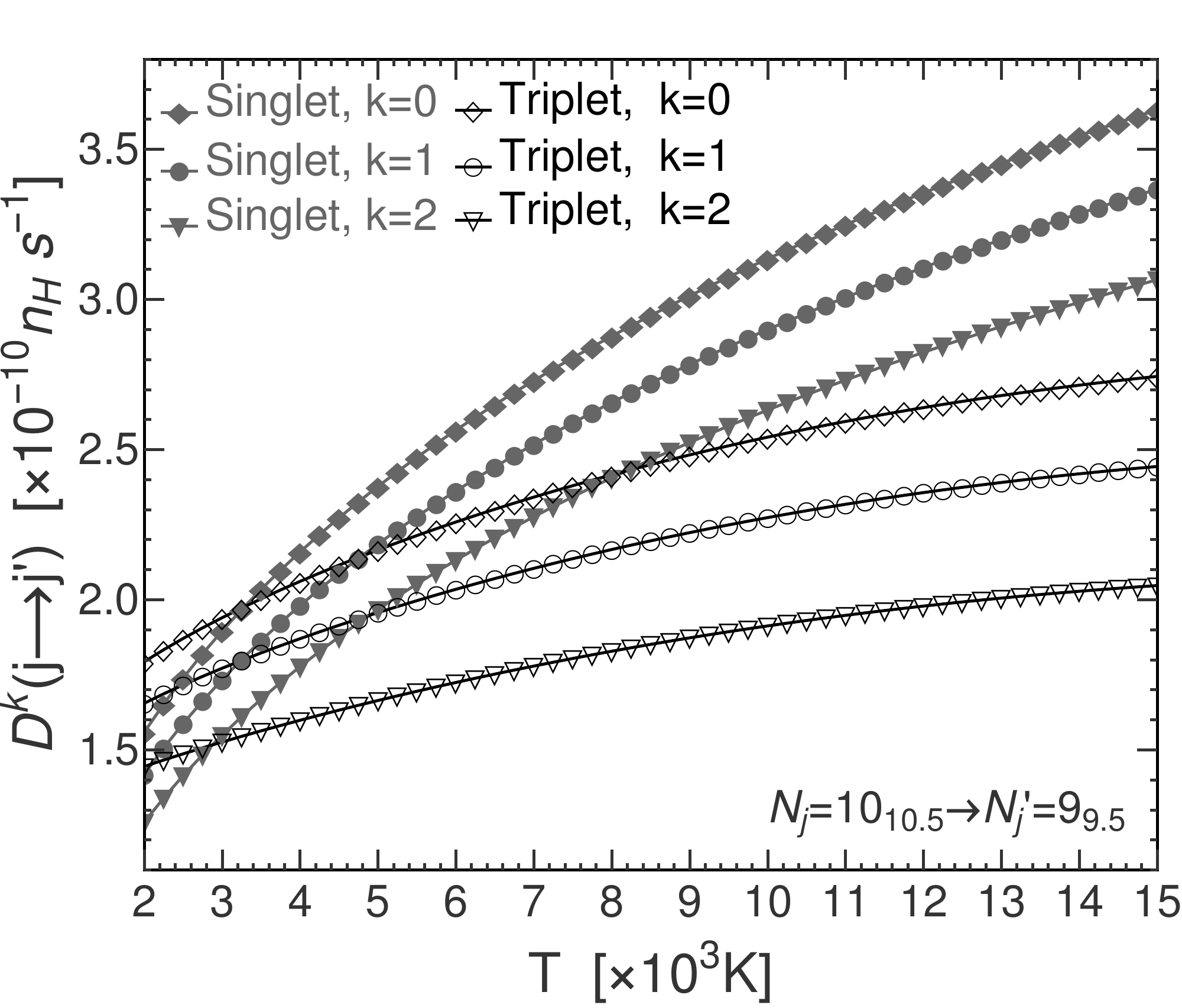} 
\includegraphics[width=8.0cm]{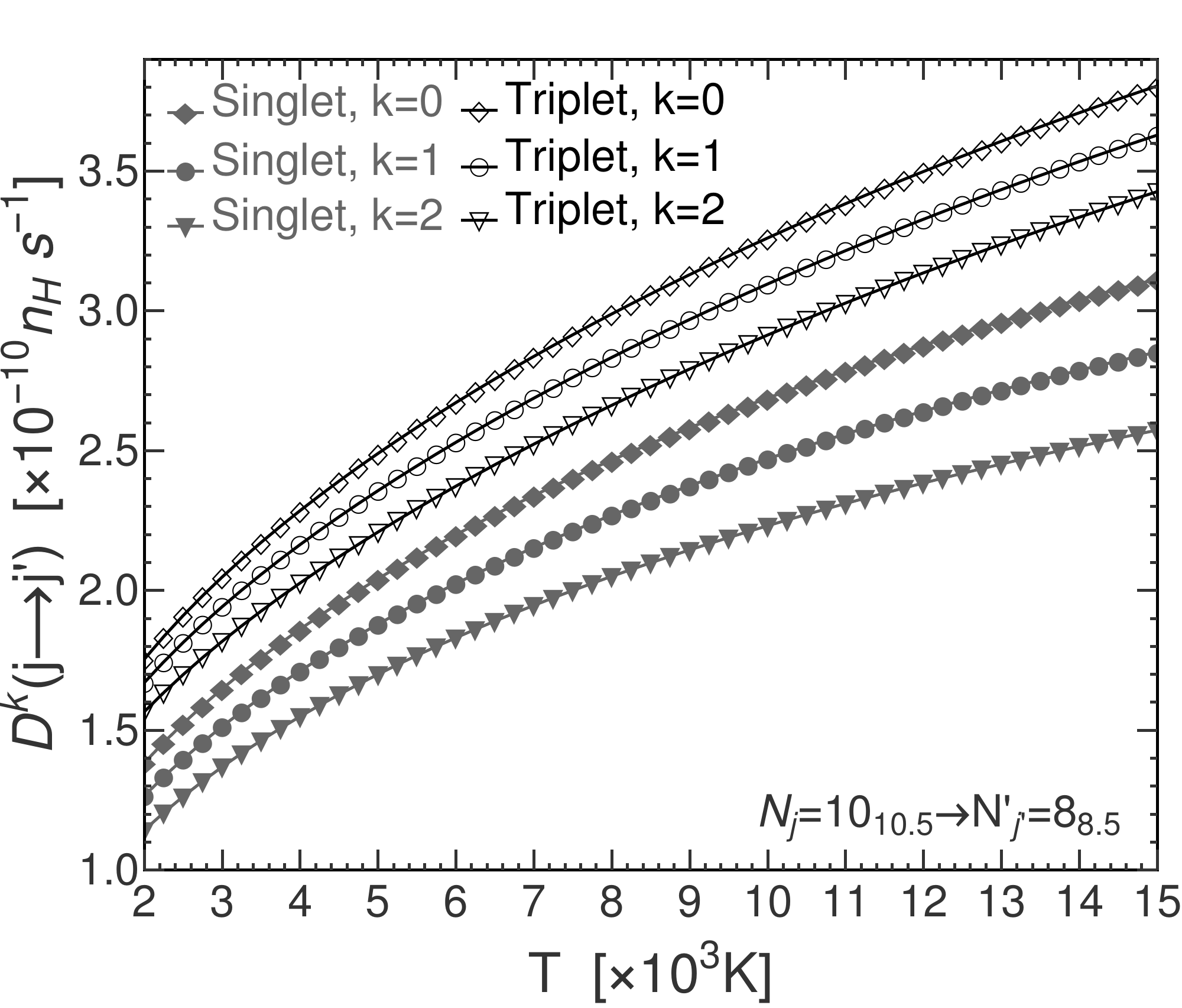}
\caption{Variation of the singlet (gray curves) and triplet (black curves) contributions to the downward transfer of polarization rates with temperature for the population, $k=0$ (diamonds), orientation, $k=1$ (circles), and alignment, $k=2$ (down-triangles), of the rotational levels $N_j = 5_{5.5}$ (upper panels) and $N_j = 10_{10.5}$ (lower panels).}
\label{fig:rate_T_deltj1}
\end{figure}

Let us now consider the thermal average of the transfer of polarization cross-sections, i.e. the rates of transfer of polarization, to study their dependence on temperature and angular momentum of the rotational levels under consideration. In Figure~\ref{fig:rate_T_deltj1}, we show the rates of transfer of polarization as functions  of temperature for the   levels $N_j = 5_{5.5}$ (upper panels) and $N_j = 10_{10.5}$ (lower panels) in the temperature range $T=2000-15000$~K.   It is clear that
in the temperature range considered,
all rates monotonically increase with increasing temperature.

%%%%%%%%%%%%%%%%%%%%%%%%%%%%%%%%%%%%%%%%%%%%
%%%%%%%%%%%%%%%%%%%%%%%%%%%%%%%%%%%%%%%%%%%%

%%%%%%%%%%%%%%%%%%%%%%%%%%%%%%%%%%%%%%%%%%%%
\begin{figure}
\centering
%\hspace{-0.7cm}
\includegraphics[width=8.0cm]{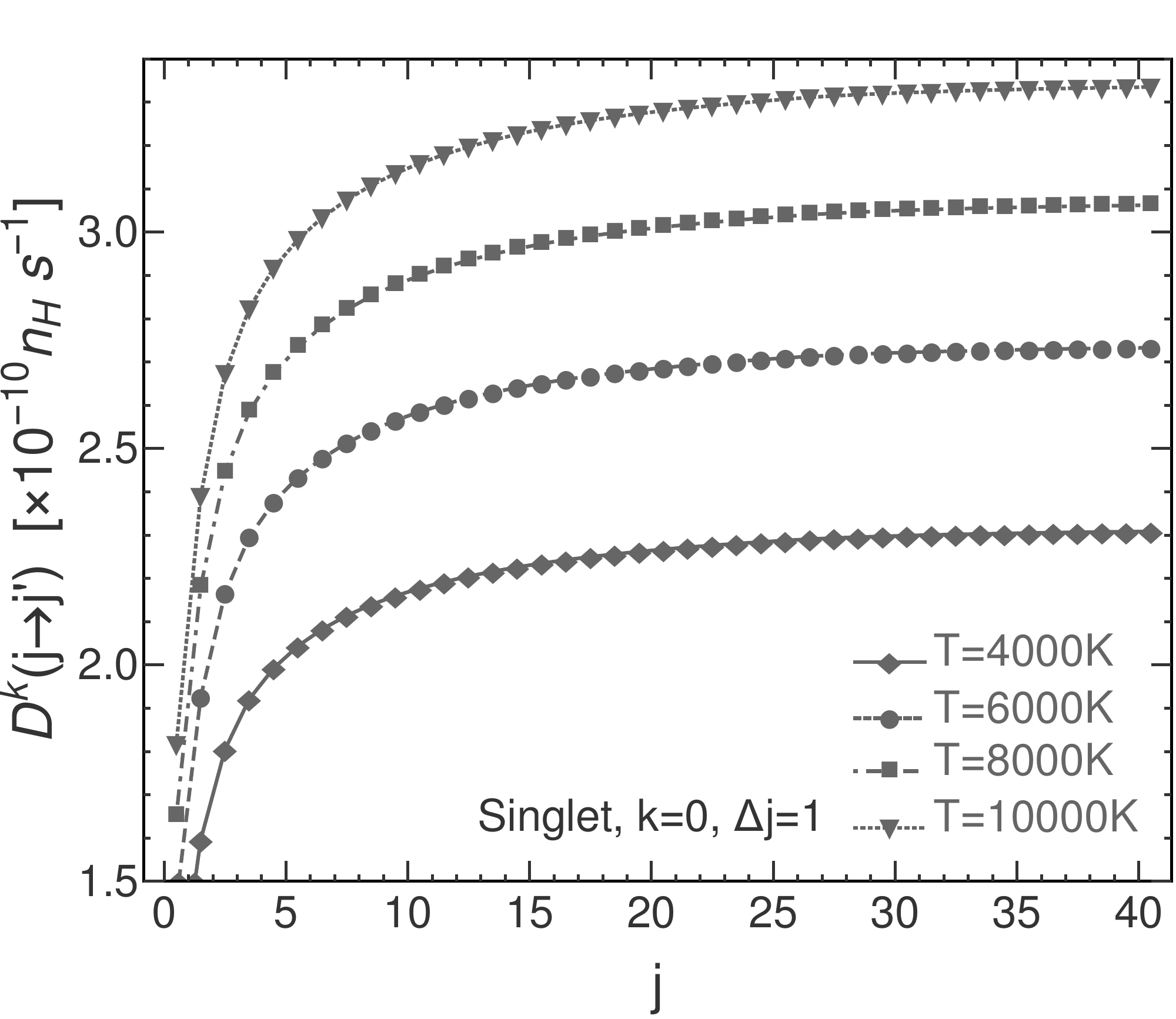}
\includegraphics[width=8.0cm]{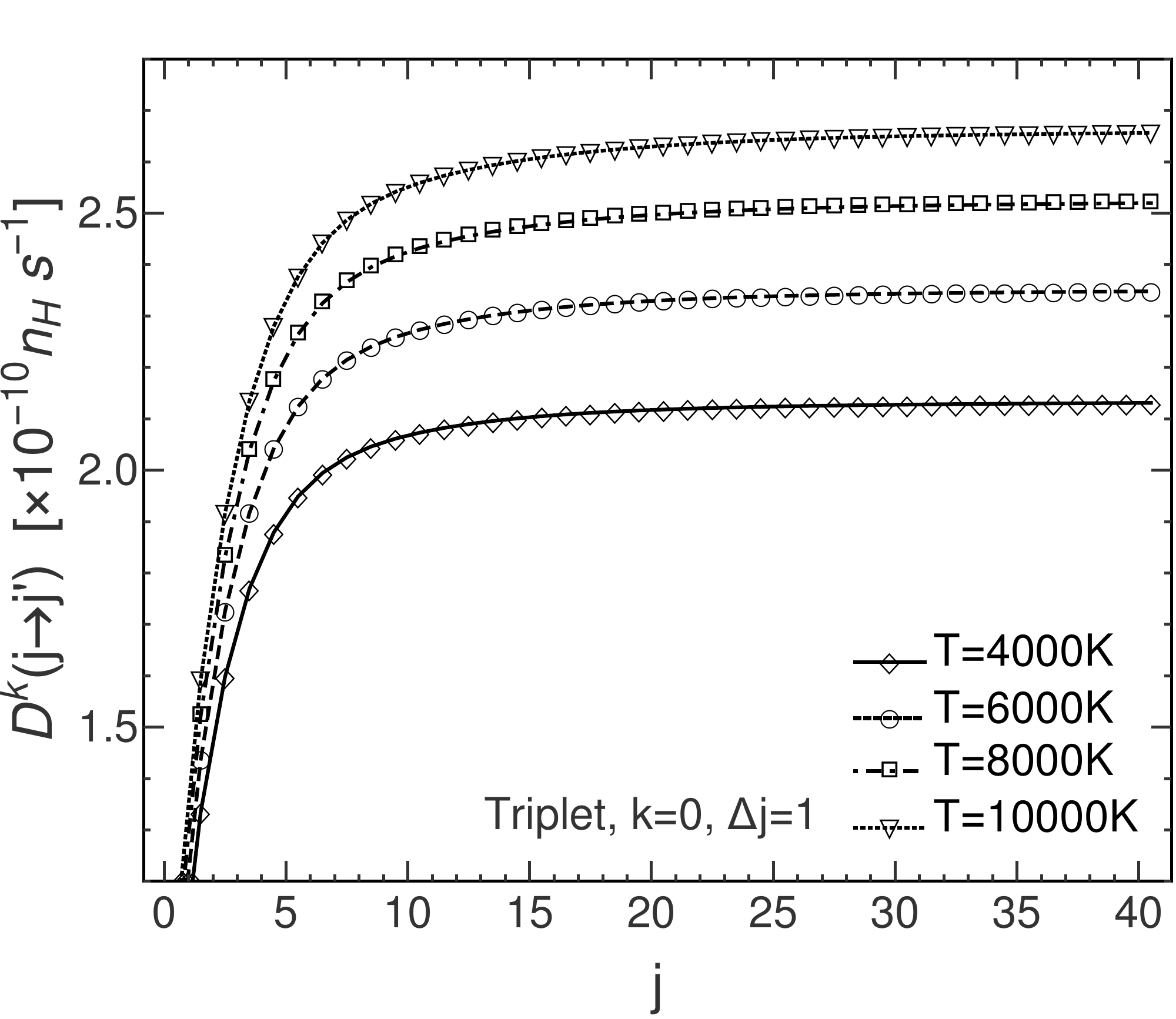} \\
\includegraphics[width=8.0cm]{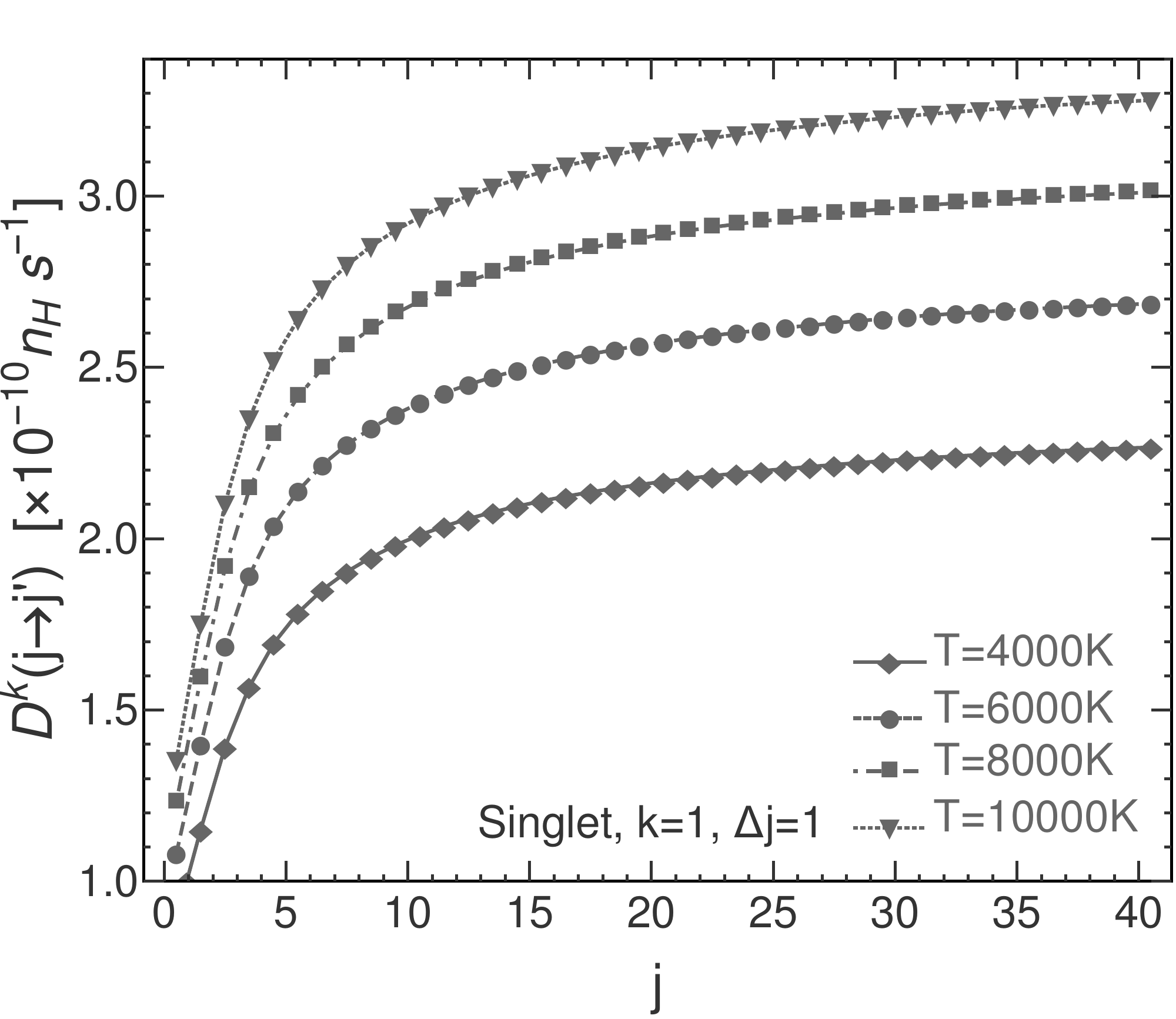}
\includegraphics[width=8.0cm]{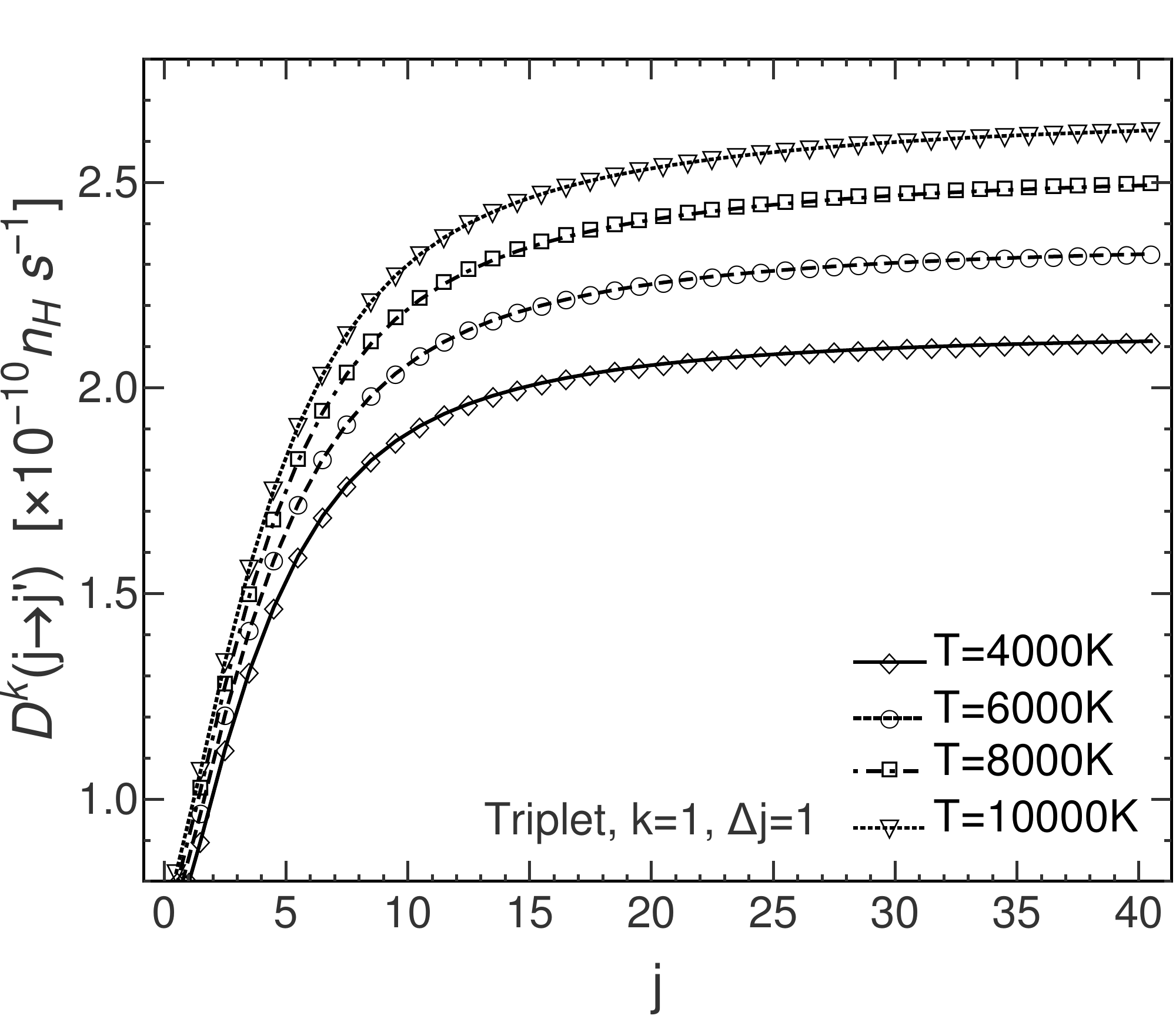}  \\
\includegraphics[width=8.0cm]{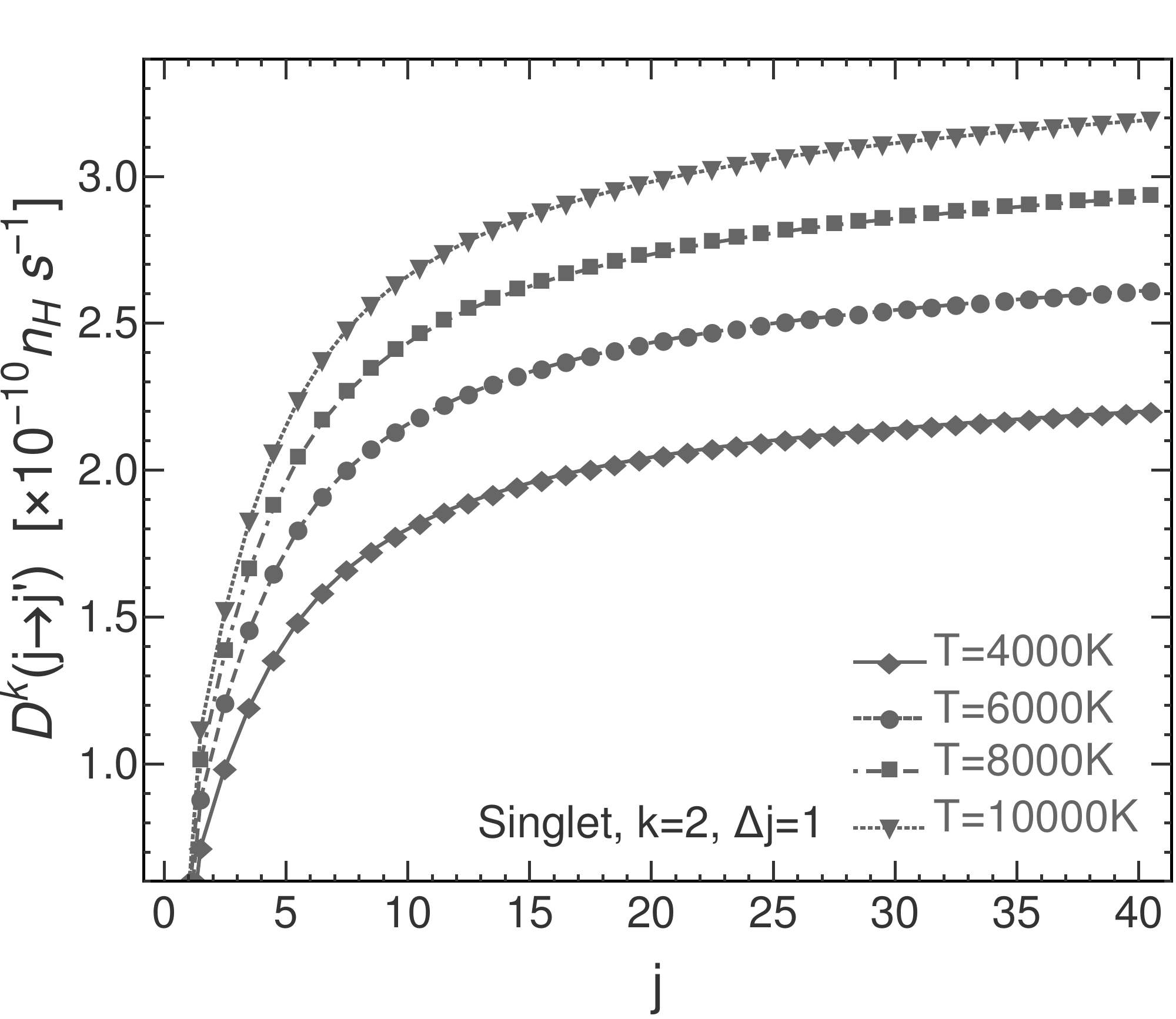}  
\includegraphics[width=8.15cm]{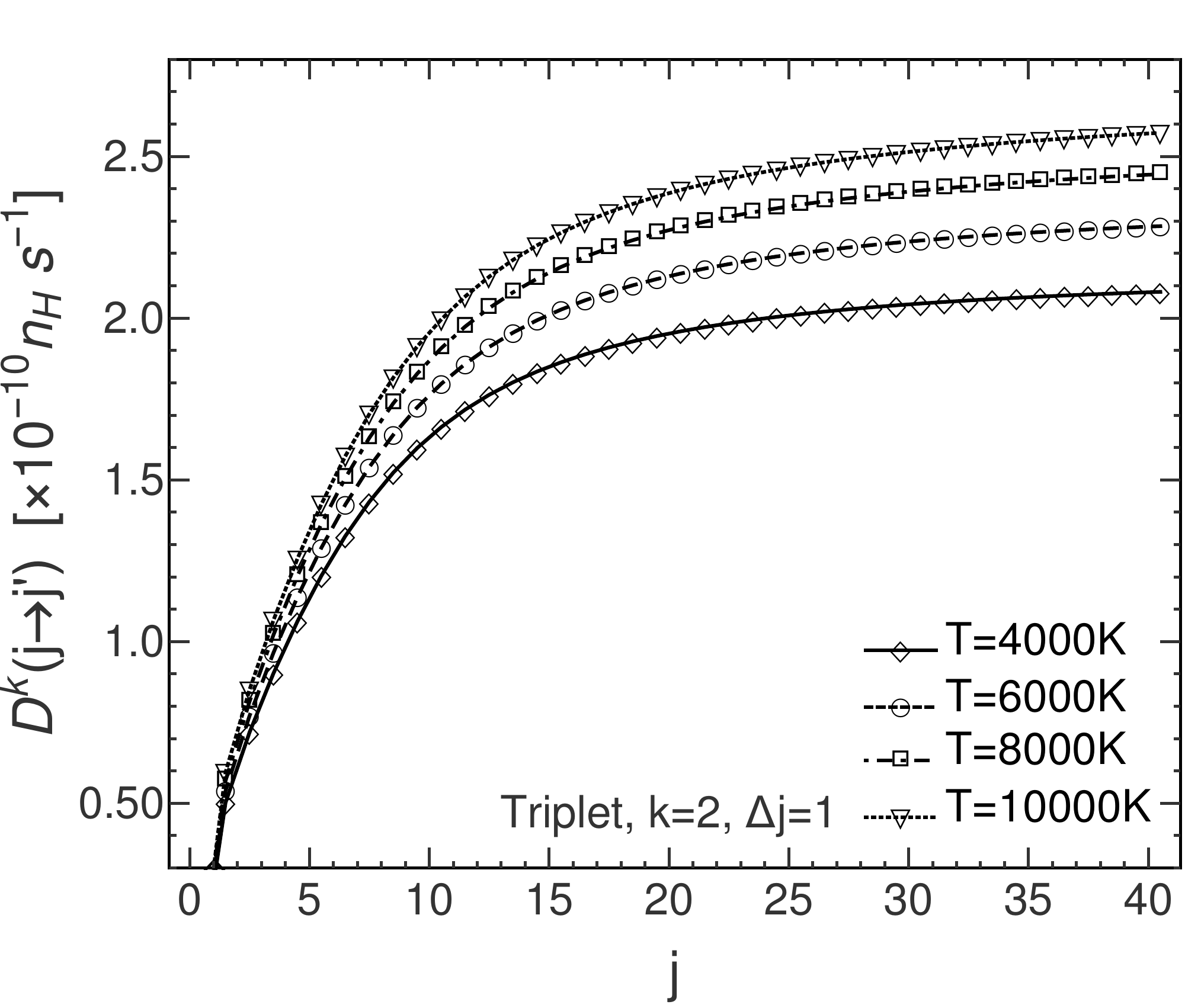}
\caption{Variation of the upward ($\Delta j = 1$) transfer of polarization rates with $j$ for the singlet (left panels) and triplet (right panels) parts of the potential.}
\label{fig:rate_j_s_k0}
\end{figure}
%%%%%%%%%%%%%%%%%%%%%%%%%%%%%%%%%%%%%%%%%%%%

In Figure~\ref{fig:rate_j_s_k0}, we show the dependence of the upward transfer of polarization rates on the angular momentum of the rotational levels considered, $j$, for various temperatures and $\Delta j =1$. The left panels show the singlet contribution,
while the right panels show the triplet contribution. It can be seen that all rates increase as $j$ increases. We note here that the behavior of the de-excitation (downward) transfer rates from a given level are similar to the upward transfer rates from the same level albiet being a bit lower (see Figure~\ref{fig:sigma_deltajm1} and the  discussion relating to it above).

\begin{figure}
\centering
%\hspace{-0.7cm}
\includegraphics[width=8.0cm]{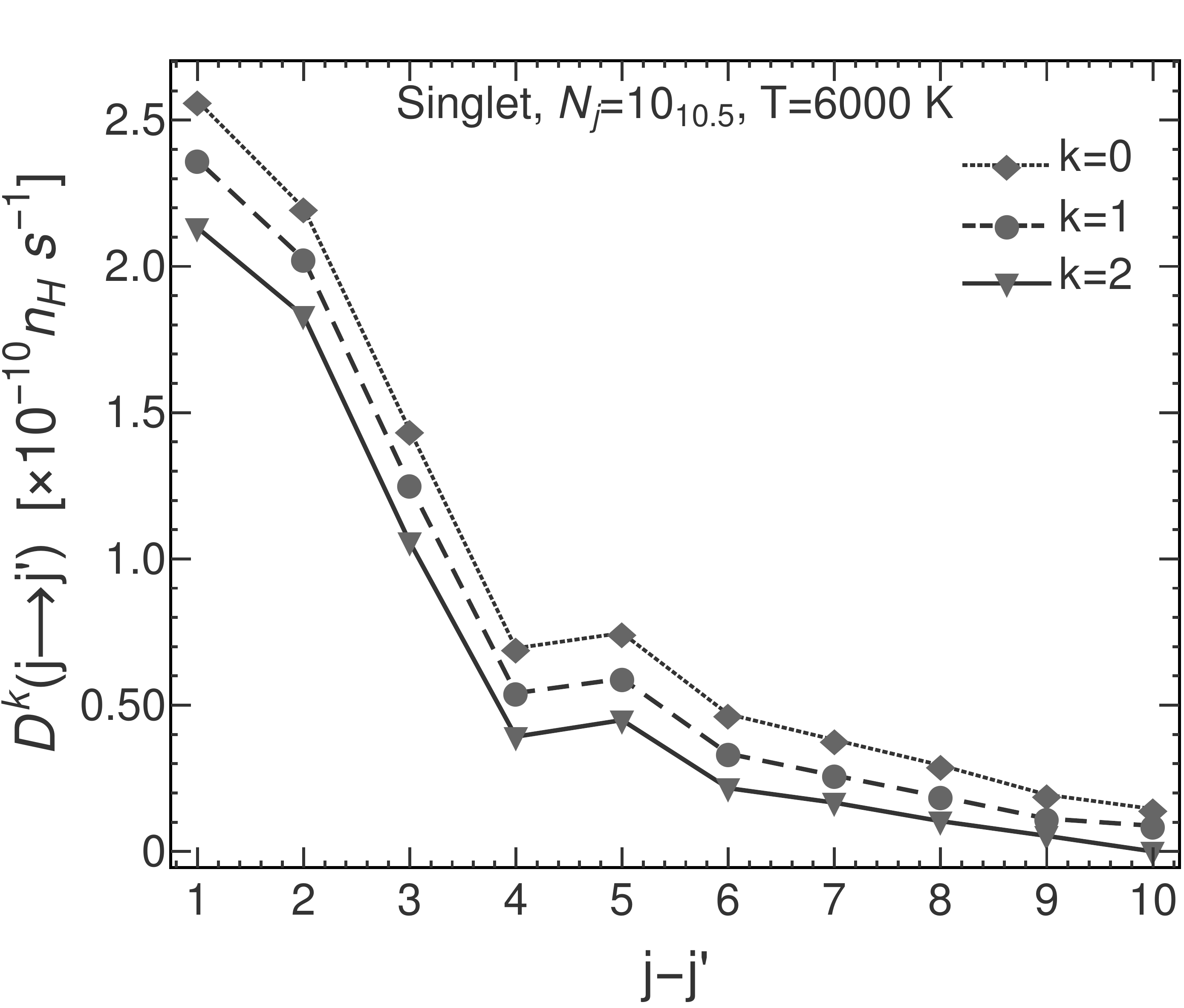} 
\includegraphics[width=8.0cm]{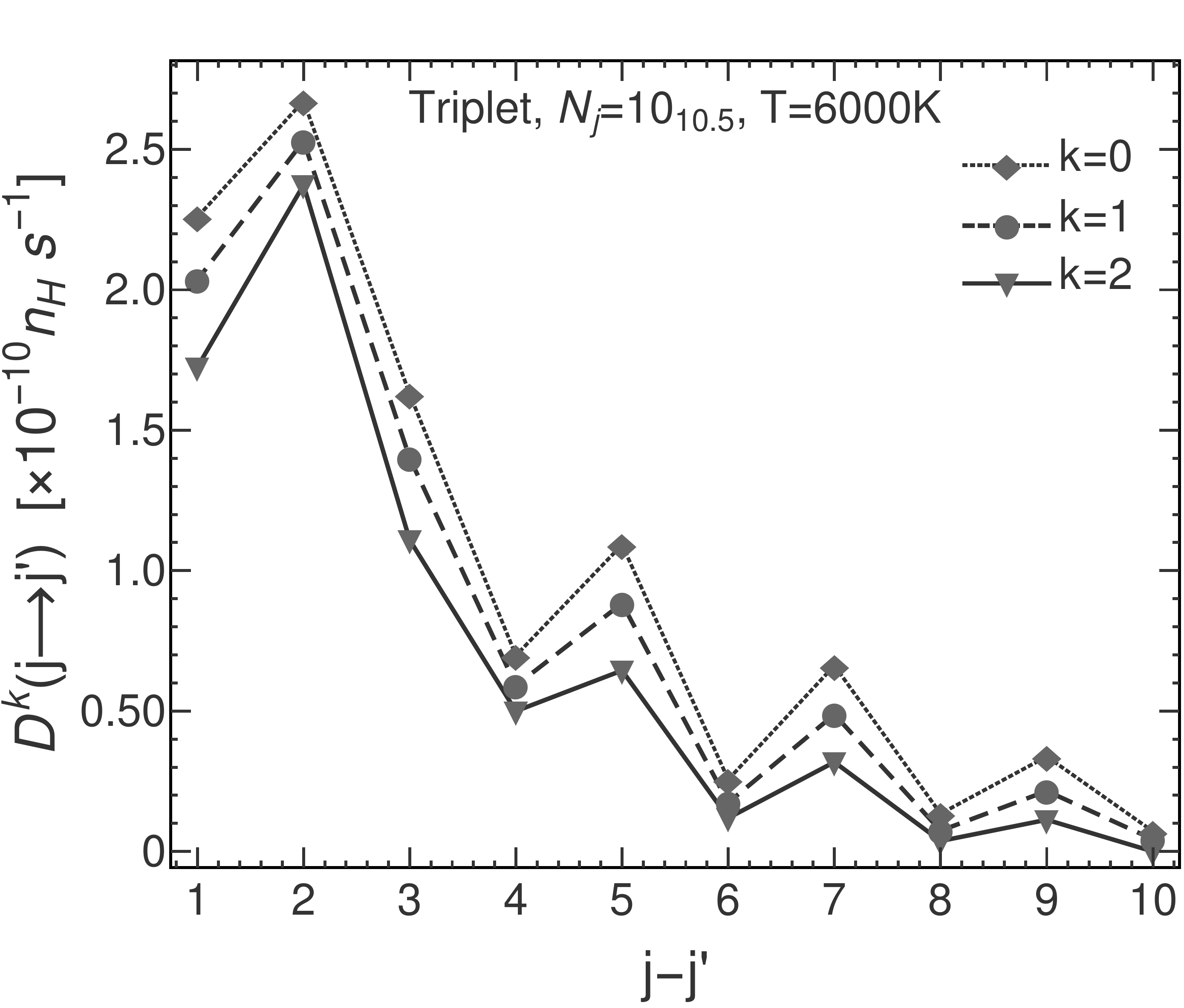}
\caption{The transfer of polarization rate of the level with $N_{j}=10_{10.5}$ as a function of $\Delta j=j-j'$ for the  singlet (left panel) and the triplet (right panel) parts of the potential, calculated at temperature $T=6000K$.}
\label{fig:rate_deltaj_s}
\end{figure}

Figure~\ref{fig:rate_deltaj_s} shows the variation with $\Delta j$ of the singlet (left panel) and triplet (right panel) contributions to the downward transfer of polarization rates for the  level $N_{j}=10_{10.5}$ at temperature $T=6000$~K. As expected, the   rates decrease  with increasing $\Delta j$. Further, the  excitation rates are expected to have behavior with $\vert \Delta j \vert$ roughly similar to that of the downward rates. We note here that the singlet and triplet contributions to the transfer of polarization rates are not very different from each other.

\subsection{Depolarization cross-sections and rates} \label{sec:depol_rate}

%%%%%%%%%%%%%%%%%%%%%%%%%%%%%%%%%%%%%%%%%%%%
\begin{figure}
\centering
%\hspace{-0.7cm}
\includegraphics[width=8.0cm]{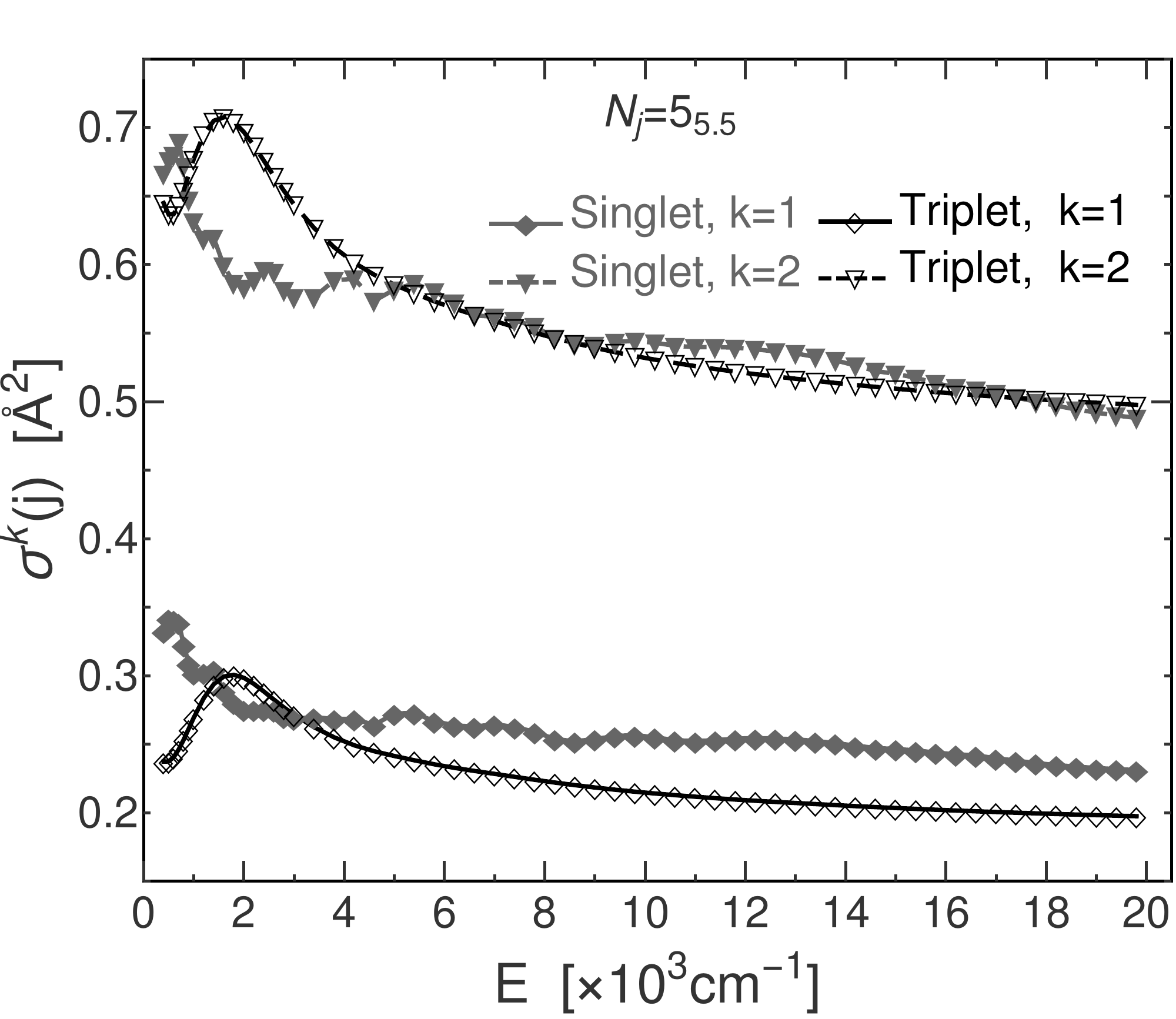} 
\includegraphics[width=8.0cm]{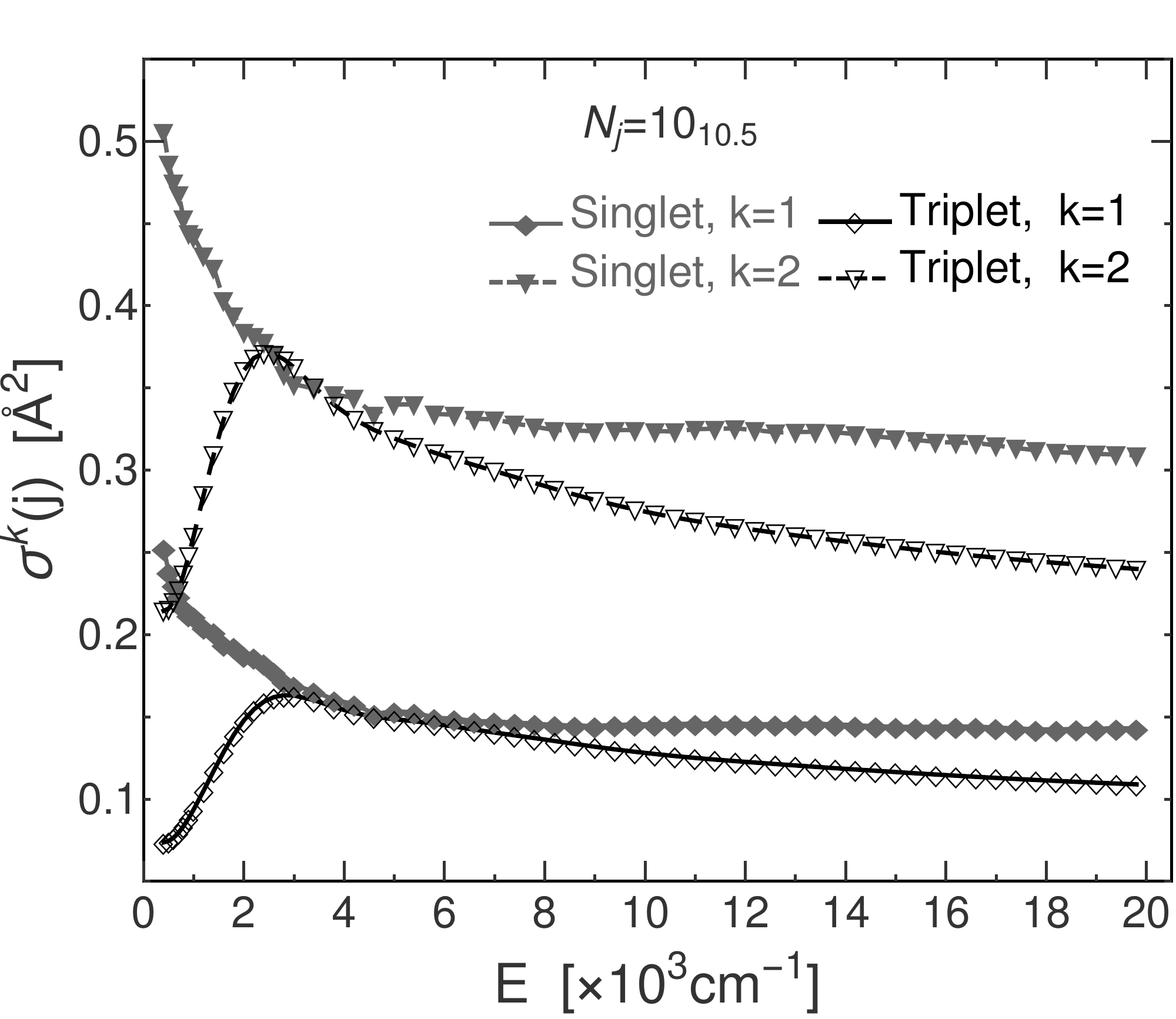}
\caption{Variation with energy of the depolarization cross-sections of the orientation, $k=1$ (diamonds), and alignment, $k=2$ (down-triangles), for the rotational levels $N_j=5_{5.5}$ (left panel) and $N_j=10_{10.5}$ (right panel). The singlet and triplet contributions are respectively represented by the gray and black curves.}
\label{fig:sigma_depol_E}
\end{figure}
%%%%%%%%%%%%%%%%%%%%%%%%%%%%%%%%%%%%%%%%%%%%

%%%%%%%%%%%%%%%%%%%%%%%%%%%%%%%%%%%%%%%%%%%% 
\begin{figure}
\centering
%\hspace{-0.7cm}
\includegraphics[width=8.0cm]{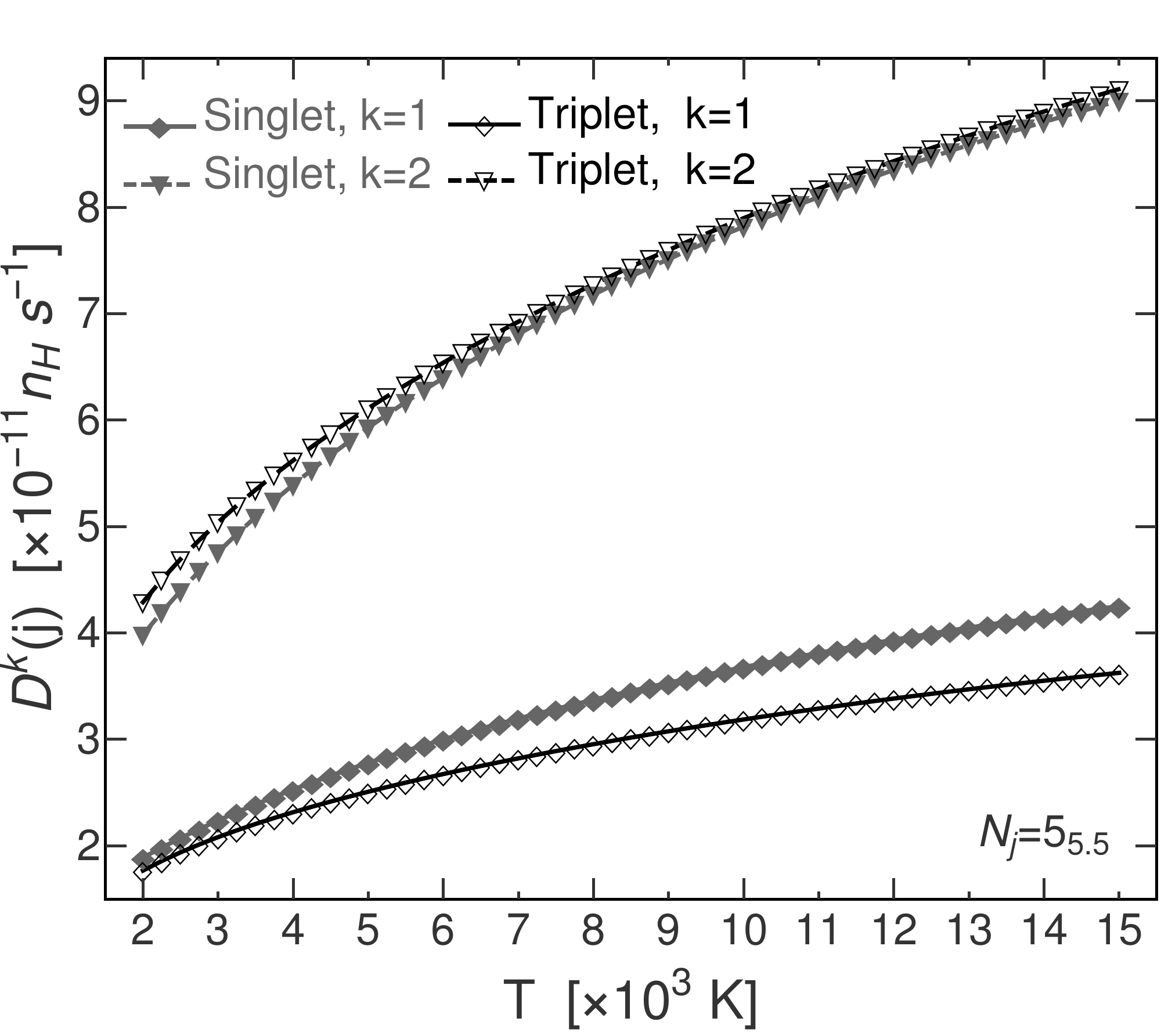} 
\includegraphics[width=8.0cm]{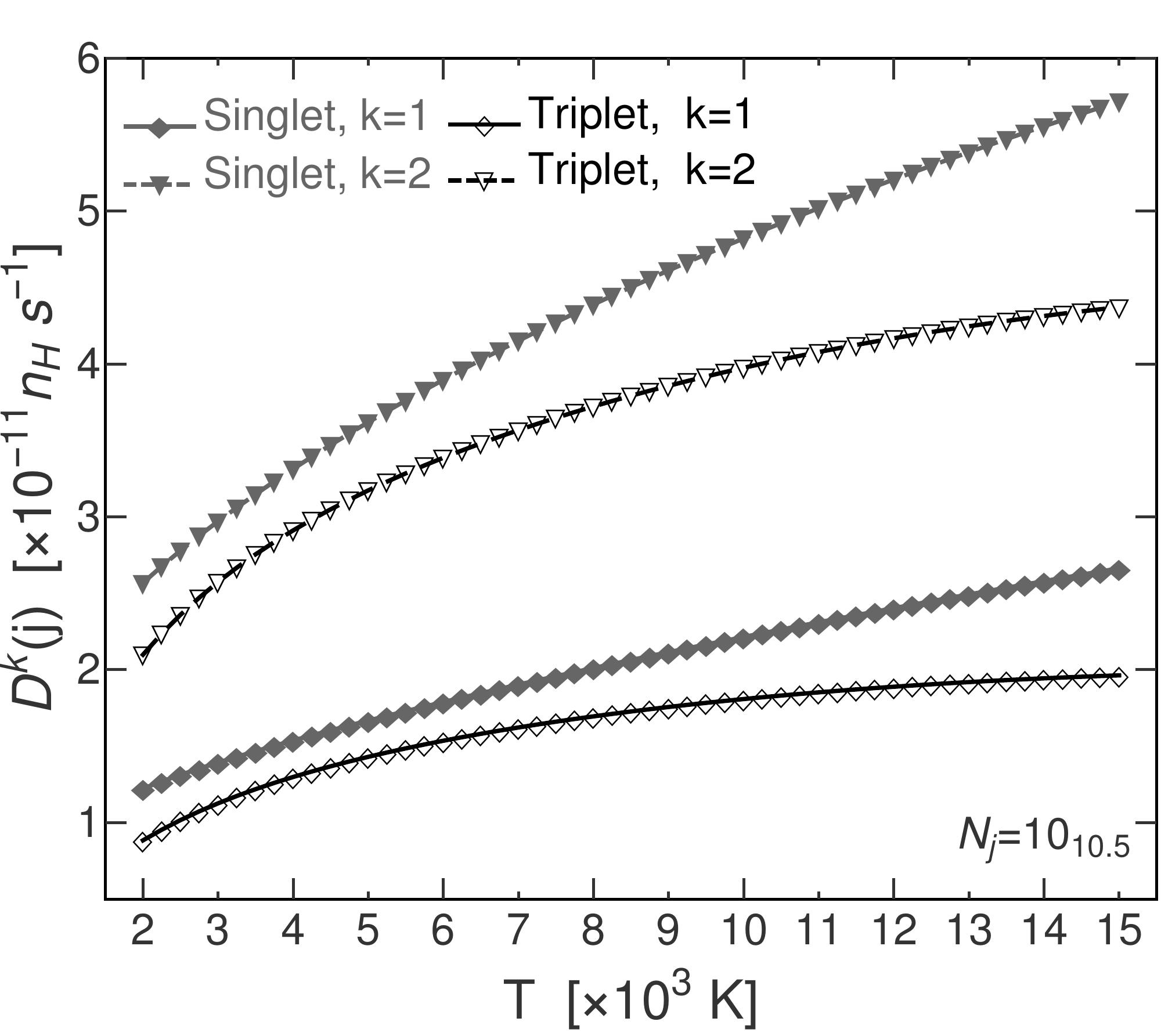}
\caption{The temperature dependence of the rates of destruction of circular, $k=1$ (diamonds), and linear, $k=2$ (down-triangles), polarization for the rotational levels $N_j=5_{5.5}$ (left panel) and $N_j=10_{10.5}$ (right panel). The singlet and triplet contributions are respectively represented by the gray and black curves.}
\label{fig:D_depol_T}
\end{figure}
%%%%%%%%%%%%%%%%%%%%%%%%%%%%%%%%%%%%%%%%%%%%

%%%%%%%%%%%%%%%%%%%%%%%%%%%%%%%%%%%%%%%%%%%%
\begin{figure}
\centering
\hspace{-0.5cm}
\includegraphics[width=8.4cm]{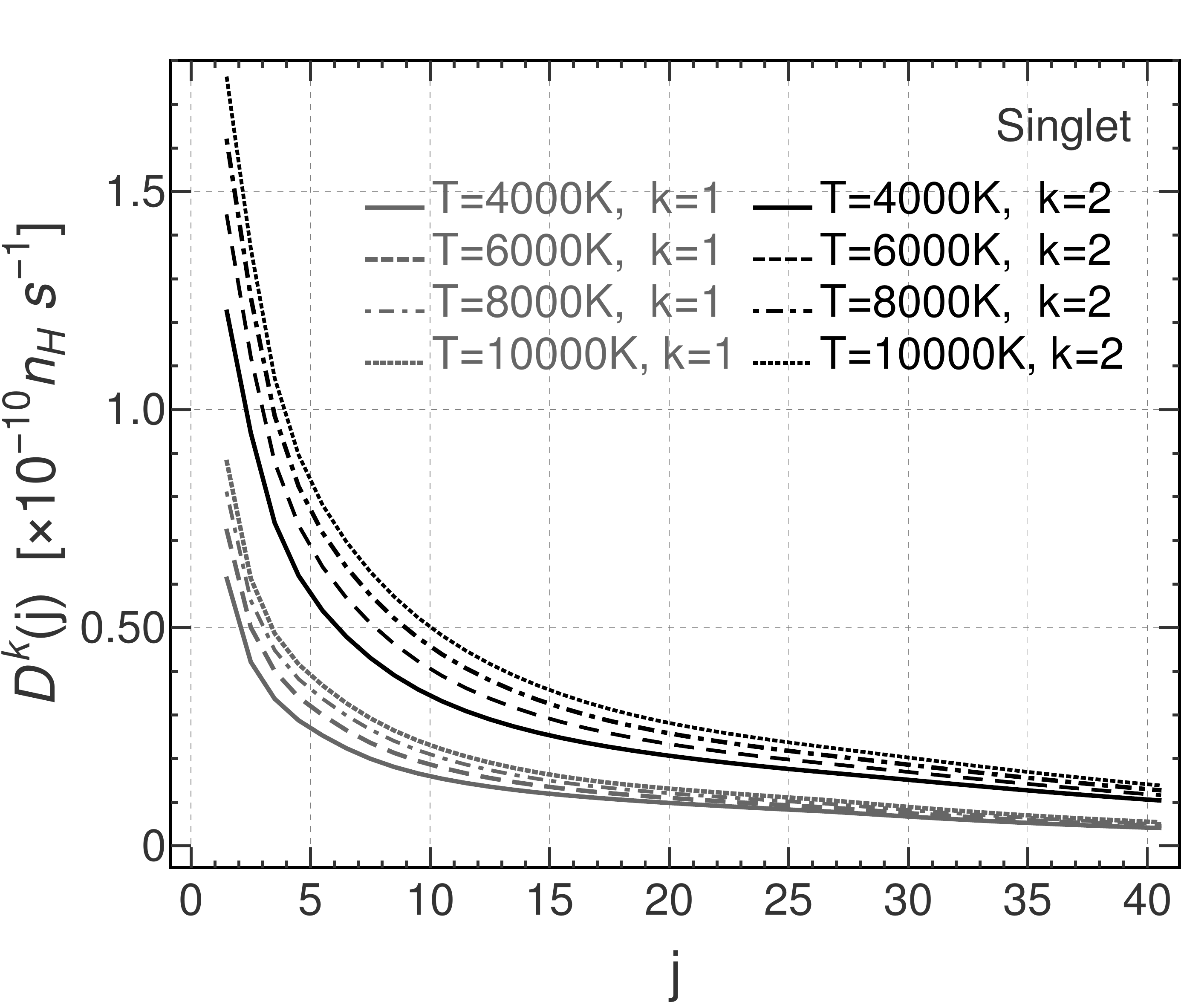} 
\includegraphics[width=8.4cm]{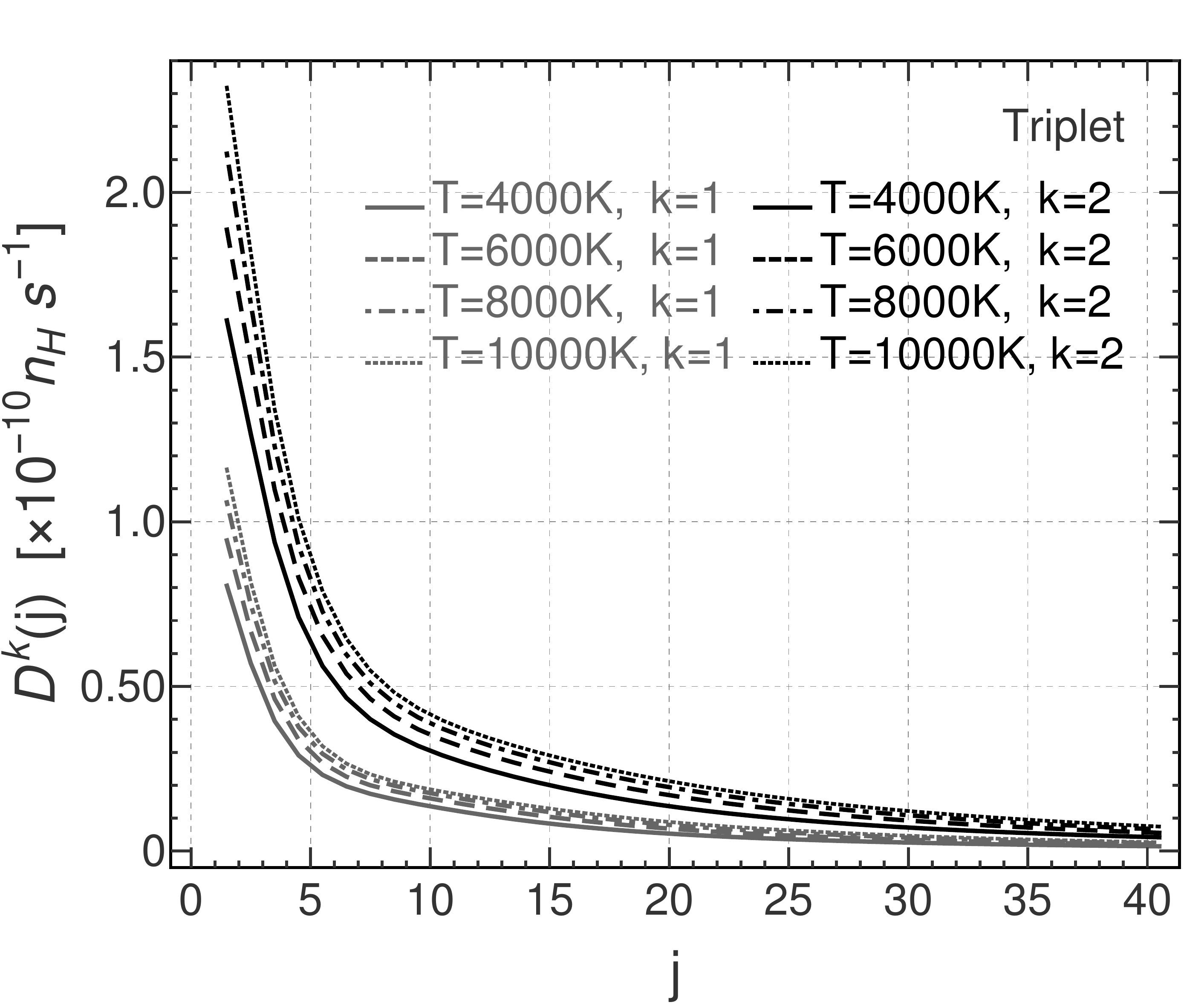} \hspace{-.5cm}
\caption{The variation with $j$ of the rates of destruction of circular, $k=1$, and linear, $k=2$, polarization of the  level $N_j$ for temperatures, $T=4000,\, 6000, \, 8000, \, 10000$~K. The singlet and triplet contributions are respectively represented by the gray and black curves.}
\label{fig:D_depol_deltaj}
\end{figure}
%%%%%%%%%%%%%%%%%%%%%%%%%%%%%%%%%%%%%%%%%%%%

Let us now turn our attention to the depolarization rates which quantify the destruction of the atomic polarization of a particular  level due to purely elastic collisions, i.e. collisions happening inside a given rotational level.
Figure~\ref{fig:sigma_depol_E}
shows  the depolarization cross-sections for the orientation, $k=1$ (diamonds), and the allignment, $k=2$ (down-triangles), of the levels $N_j= 5_{5.5}$ (left panel) and $N_j=10_{10.5}$ (right panel).  One can remark that the cross-sections for the destruction of linear polarization, $\sigma^{k=2} (j)$, is roughly twice as large as that for circular polarization $\sigma^{k=1} (j)$.
We next show the dependence of the depolarization rates on temperature in Figure~\ref{fig:D_depol_T} for the  levels $N_j=5_{5.5}$ (left panel) and $N_j=10_{10.5}$ (right panel). As expected, all the rates increase  as the temperature increases.

In Figure~\ref{fig:D_depol_deltaj}, we show the variation of the depolarization rates with the angular momentum, $j$, of the  levels under consideration for various temperatures. The gray curves represent the singlet contribution, while black curves represent the triplet contribution. 
Unlike 
the transfer of polarization rates, the depolarization rates decrease with increasing $j$. Moreover, the linear depolarization rates are roughly twice as high as the circular depolarization rates. It is interesting to note here that the contribution of the triplet component of the potential to the depolarization rates is larger than that of the singlet component for small $j$ whereas for levels with large $j$ values the singlet contribution is a bit larger than the triplet contribution. It is also interesting to remark that the depolarization rates for different temperatures converge as $j$ increases.

In Table~\ref{tab:D_depol_j}, we present fits of the singlet and triplet contribution to the destruction of circular ($k=1$) and linear ($k=2$) polarization rates for various temperatures. 
%These variation laws are accurate within roughly 10\% up to $j=40$. 
These variation laws give results with precision better than 10\% up to $j=40.5$. Our results could be   implemented   in numerical codes concerned with the  simulations of the scattering polarization to obtain accurately the magnetic field in the quiet Sun.  
We note here
that both the singlet and triplet contributions to the depolarization rates exhibit a very mild oscillatory behavior on top of their decreasing behavior with increasing $j$. However, as shown in Table~\ref{tab:D_depol_j}, the behavior of all the singlet and some of the triplet depolarization rates can be described by simple power laws within the intended accuracy.

%we obtain  useful variation  laws of the polarization transfer rates with the temperature and the angular momentum $j$. 

%Finally, by using sophisticated numerical,  it is found that the behavior of the depolarization   rates  obey non linear variation laws with  the angular momentum $j$. These laws would allow  quick calculations of the depolarizing collisional rates  and can be implemented in radiative transfer codes to synthetize polarization profiles (see Table 2). 

%%%%%%%%%%%%%%%%%%%%%%%%%%%%%%%%%%%%%%%%%%%%
\begin{table}
\centering
\caption{Depolarization rates ($\times 10^{-10} n_H \ s^{-1}$) as functions of $j$.}
\begin{tabular}{|c|c|c|c|} 
\hline
\hline
T (K) & $k$ &  $D^k(j)$ [Singlet] &  $D^k(j)$ [Triplet]  \\
\hline
2000 & 1 &
$\frac{0.580672}{j^{0.659399}} \!-\! \frac{j^{5.34649}}{4.664955 \!\times\! 10^{11}}$  &
$\frac{1.93722 \sin (0.716361 j)}{j^{3.05797}} \!+\! \frac{98.9513 \sin (0.034423 j)}{j^{2.54382}} \!-\! \frac{15.2373 \sin (0.285721 j)}{j^{3.25848}}$\\ 
     & 2 &
$\frac{1.20891}{j^{0.651866}} \!-\! \frac{j^{5.45485}}{3.564565 \!\times\! {10}^{11}}$ &
$ \frac{105.145 \sin (0.0421442 j)}{j^{2.28329}} \!-\! \frac{2.11805 \sin (1.08345 j)}{j^{3.43576}} \!-\! \frac{j^{28.7795} \sin (0.763456 j)}{1.558629 \!\times\! 10^{49}} \!-\! \frac{5.8483 \sin (0.29407 j)}{j^{2.63312}}$\\ 
\hline
3000 & 1 &
$\frac{0.709969}{j^{0.686909}} \!-\! \frac{j^{4.42095}}{1.595641 \!\times\! 10^{9}}$ &
$\frac{15213.3}{j^{0.29214}} \!-\! \frac{27200.8}{j^{0.290325}} \!+\! \frac{11988.5}{j^{0.28806}}$ \\
     & 2 &
$\frac{952.546}{j^{0.433596}} \!-\! \frac{951.134}{j^{0.433343}}$ &
$ \frac{52.7609 \sin (0.053414 j)}{j^{1.99877}} \!+\! \frac{0.10963 \sin (0.589314 j)}{j^{1.44464}} \!-\!  \frac{j^{30.3895} \sin (0.767695 j)}{2.862860 \!\times\! 10^{51}} \!-\! \frac{1.94657 \sin (1.1667 j)}{j^{3.4203}} $\\
\hline
4000 & 1 &
$\frac{0.81136}{j^{0.699613}} \!-\!  \frac{j^{3.42274}}{6.966464 \!\times\! 10^{8}}$ &
$\frac{21010.6}{j^{0.339994}} \!-\! \frac{37614.6}{j^{0.338475}} \!+\! \frac{16605.2}{j^{0.336582}}$ \\
     & 2 &
$\frac{1309.16}{j^{0.414439}} \!-\! \frac{1307.57}{j^{0.414214}}$ &
$\frac{68.5569 \sin (0.0463297 j)}{j^{2.00335}} \!-\! \frac{1.80188 \sin (1.25799 j)}{j^{3.02838}}$\\ 
\hline
5000 & 1 &
$\frac{0.892242}{j^{0.701794}} \!-\!  \frac{j^{2.31421}}{4.513851 \!\times\! 10^{6}}$ &
$\frac{893.269}{j^{0.584877}} \!-\! \frac{892.011}{j^{0.584538}}$ \\

     & 2 &
$\frac{1613.83}{j^{0.401726}} \!-\! \frac{1612.08}{j^{0.401517}}$ &
$\frac{80.6866 \sin (0.0425921 j)}{j^{2.00046}} \!-\! \frac{1.8668 \sin (1.27448 j)}{j^{2.94507}}$\\
\hline
6000 & 1 &
$\frac{0.957088}{j^{0.694838}} \!-\! \frac{j^{1.41925}}{1.37194 \!\times\! 10^{4}}$ &
$\frac{1111.39}{j^{0.596797}} \!-\! \frac{1110.04}{j^{0.59651}}$ \\
     & 2 &
$\frac{1629.84}{j^{0.392947}} \!-\! \frac{1627.96}{j^{0.39272}}$ &
$\frac{94.2274 \sin (0.0391989 j)}{j^{2.00339}} \!-\! \frac{1.97187 \sin (1.27964 j)}{j^{2.90373}}$\\
\hline
7000 & 1 &
$\frac{1.01366}{j^{0.685892}} \!-\! \frac{j^{0.999696}}{7.41299 \!\times\! 10^{4}}$ &
$\frac{36.7084 \sin (0.0453646 j)}{j^{1.99647}} \!-\! \frac{0.412163 \sin (1.60238 j)}{j^{2.30586}}$ \\
     & 2 &
$\frac{1962.72}{j^{0.386759}} \!-\! \frac{1960.73}{j^{0.386556}}$ &
$\frac{109.44 \sin (0.0360508 j)}{j^{2.00927}} \!-\! \frac{2.09951 \sin (1.27903 j)}{j^{2.88506}}$\\
\hline
8000 & 1 &
$\frac{1.06554}{j^{0.679555}} \!-\!  \frac{j^{0.838967}}{1.50584 \!\times\! 10^{3}}$ &
$\frac{41.3965 \sin (0.0426049 j)}{j^{2.0026}} \!-\! \frac{0.435909 \sin (1.59865 j)}{j^{2.29474}}$
\\
     & 2 &
$\frac{2060.22}{j^{0.382413}} \!-\! \frac{2058.13}{j^{0.382207}}$  &
$\frac{126.848 \sin (0.0330478 j)}{j^{2.01682}} \!-\! \frac{2.24232 \sin (1.27506 j)}{j^{2.8798}}$\\
\hline
9000 & 1 &
$\frac{1.11331}{j^{0.676387}} \!-\!  \frac{j^{0.804209}} {1.85496 \!\times\! 10^{3}}$ &
$\frac{46.5176 \sin (0.0399658 j)}{j^{2.0099}} \!-\! \frac{0.462707 \sin (1.59339 j)}{j^{2.29394}}$ \\
     & 2 &
$\frac{2117.99}{j^{0.379395}} \!-\! \frac{2115.8}{j^{0.379184}}$ &
$\frac{147.34 \sin (0.030095 j)}{j^{2.02545}} \!-\! \frac{2.39696 \sin (1.26892 j)}{j^{2.88306}}$\\
\hline
10000 & 1 &
$\frac{1.15734}{j^{0.675571}} \!-\!  \frac{j^{0.841397}} {1.70383 \!\times\! 10^{3}}$ &
$\frac{172.447 \sin (0.0270968 j)}{j^{2.03482}} \!-\! \frac{2.56193 \sin (1.26123 j)}{j^{2.89201}}$ \\
      & 2 &
$\frac{2407.5}{j^{0.37736}} \!-\! \frac{2405.22}{j^{0.377167}}$ &
$\frac{148.979 \sin (0.0300641 j)}{j^{2.01433}} \!-\! \frac{1.30349 \sin (1.70308 j)}{j^{2.77411}} \!-\! \frac{0.591264 \sin (1.26576 j)}{j^{1.8358}} \!-\! \frac{0.262735 \sin (0.985727 j)}{j^{1.3754}}$\\
\hline
11000 & 1 &
$\frac{1.19809}{j^{0.676055}} \!-\! \frac{j^{0.925462}} {1.29377 \!\times\! 10^{3}}$ &
$\frac{2.03058 \sin (0.575704 j)}{j^{2.47863}} \!+\! \frac{25.1946 \sin (0.0549002 j)}{j^{1.82949}} \!+\!  \frac{j^{27.0852} \sin (0.689243 j)}{4.025176 \!\times\! 10^{46}} \!-\! \frac{1.56857 \sin (1.22555 j)}{j^{3.63018}} $ \\
      & 2 &
$\frac{2129.78}{j^{0.376117}} \!-\! \frac{2127.42}{j^{0.375892}}$ &
$\frac{6.24505 \sin (0.444197 j)}{j^{2.6285}} \!+\! \frac{79.9193 \sin (0.0429266 j)}{j^{1.87806}} \!-\! \frac{5.34946 \sin (0.979811 j)}{j^{3.43624}} $\\
\hline
12000 & 1 &
$\frac{1.23592}{j^{0.676985}} \!-\! \frac{j^{1.04281}} {8.61328 \!\times\! 10^{4}}$ &
$\frac{2.35808 \sin (0.562536 j)}{j^{2.52625}} \!+\! \frac{26.0198 \sin (0.0539131 j)}{j^{1.82608}} \!+\!  \frac{j^{35.9765} \sin (0.84497 j)}{2.084109 \!\times\! 10^{60}} \!-\! \frac{1.73953 \sin (1.20264 j)}{j^{3.63959}}$ \\
      & 2 &
$\frac{2269.39}{j^{0.375461}} \!-\! \frac{2266.96}{j^{0.375244}}$ &
$\frac{200.538 \sin (0.0245231 j)}{j^{2.03363}} \!-\! \frac{1.43117 \sin (1.70595 j)}{j^{2.80279}} \!-\! \frac{0.676838 \sin (1.26708 j)}{j^{1.8719}} \!-\! \frac{0.332371 \sin (0.980758 j)}{j^{1.42887}}$\\
\hline
13000 & 1 &
$\frac{1.27101}{j^{0.677788}} \!-\!  \frac{j^{1.18463}} {5.2231 \!\times\! 10^{4}}$ &
$\frac{2.81554 \sin (0.548392 j)}{j^{2.59105}} \!+\! \frac{26.398 \sin (0.0533797 j)}{j^{1.81963}} \!-\!  \frac{j^{32.1112} \sin (0.766856 j)}{3.0086945 \!\times\! 10^{54}} \!-\! \frac{1.98032 \sin (1.1719 j)}{j^{3.6541}}$ \\
      & 2 &
$\frac{2409.81}{j^{0.375299}} \!-\! \frac{2407.31}{j^{0.37509}}$ &
$\frac{82.1085 \sin (0.0420508 j)}{j^{1.86367}} \!-\! \frac{1.79062 \sin (1.21992 j)}{j^{3.34602}} \!+\! \frac{10.8506 \sin (0.416841 j)}{j^{2.84112}} \!-\! \frac{5.60116 \sin (0.856063 j)}{j^{3.29195}} $\\ 
\hline
14000 & 1 &
$\frac{1.30339}{j^{0.678139}} \!-\!  \frac{j^{1.344}} {2.96907 \!\times\! 10^{4}}$ &
$\frac{3.31323 \sin (0.533883 j)}{j^{2.64882}} \!+\! \frac{26.7466 \sin (0.0527932 j)}{j^{1.81369}} \!-\!  \frac{j^{32.483} \sin (0.766833 j)}{7.050008 \!\times\! 10^{55}} \!-\! \frac{2.22853 \sin (1.14656 j)}{j^{3.66373}}$ \\
      & 2 &
$\frac{2169.11}{j^{0.375563}} \!-\! \frac{2166.55}{j^{0.375326}}$ &
$\frac{83.361 \sin (0.0415676 j)}{j^{1.8586}} \!-\! \frac{2.02322 \sin (1.20303 j)}{j^{3.35631}} \!+\! \frac{12.9384 \sin (0.404303 j)}{j^{2.88833}} \!-\! \frac{6.31348 \sin (0.838442 j)}{j^{3.30842}} $\\
\hline
15000 & 1 &
$\frac{1.3331}{j^{0.677885}} \!-\!  \frac{j^{1.5146}} {1.62161 \!\times\! 10^{4}}$ &
$\frac{3.83097 \sin (0.519459 j)}{j^{2.69609}} \!+\! \frac{27.0655 \sin (0.0521695 j)}{j^{1.80817}} \!+\! \frac{j^{37.0213} \sin (0.844876 j)}{3.493497 \!\times\! 10^{62}} \!-\! \frac{2.47177 \sin (1.12596 j)}{j^{3.67129}}$\\
      & 2 &
$\frac{1956.09}{j^{0.376172}} \!-\! \frac{1953.47}{j^{0.375905}}$ &
$\frac{11.5399 \sin (0.393126 j)}{j^{2.75561}} \!+\! \frac{84.5611 \sin (0.0413531 j)}{j^{1.85581}}  \!+\!  \frac{j^{29.4812} \sin (0.839229 j)}{1.978907 \!\times\! 10^{50}}  \!-\! \frac{7.59474 \sin (0.935008 j)}{j^{3.46111}}$\\
\hline
\hline
\end{tabular}
\label{tab:D_depol_j}
\end{table}
%%%%%%%%%%%%%%%%%%%%%%%%%%%%%%%%%%%%%%%%%%%%

\subsection{Comparsion between the singlet and triplet contributions} \label{sec:comparison}

%%%%%%%%%%%%%%%%%%%%%%%%%%%%%%%%%%%%%%%%%%%%
\begin{figure}
\centering
%\hspace{-0.7cm}
\includegraphics[width=8.0cm]{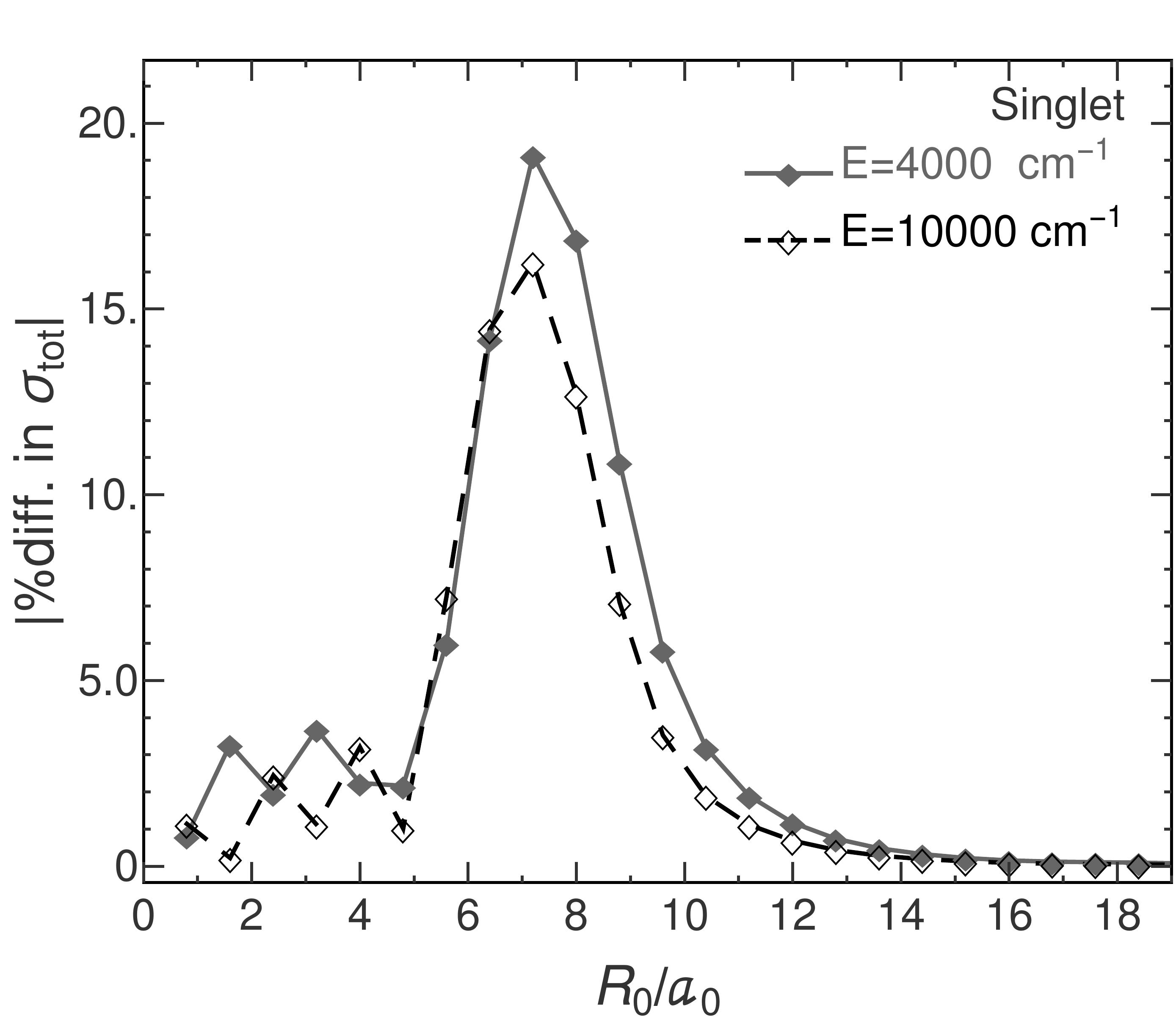} 
\includegraphics[width=8.0cm]{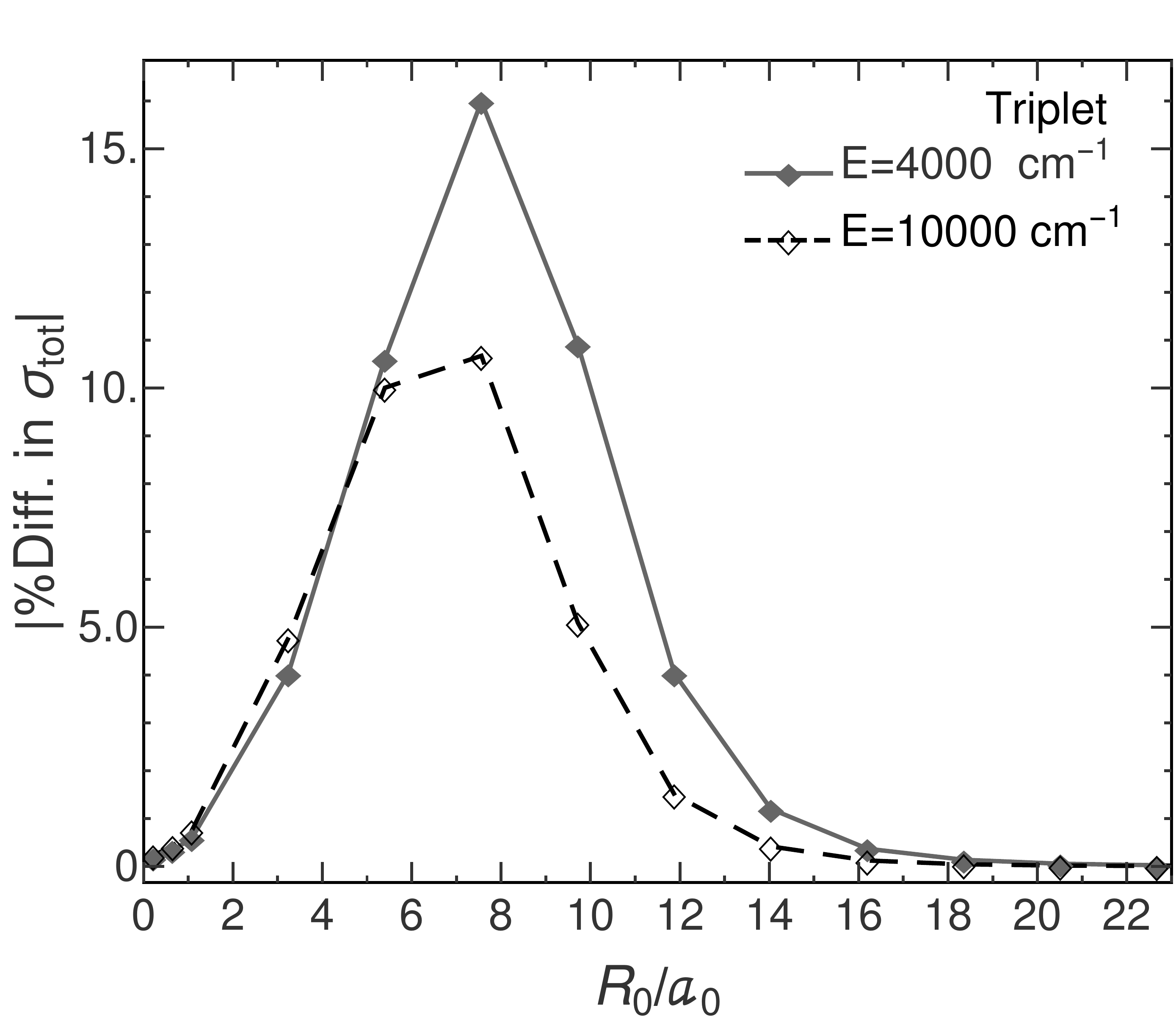}
\caption{Ratio of the total cross-sections calculated with the perturbed potential to those calculated with the unperturbed potential for collision energy $E=4000$~cm$^{-1}$ (solid gray lines) and $E=10000$~cm$^{-1}$ (dashed black lines).}
\label{fig:sigma_s_perturbed}
\end{figure}
%%%%%%%%%%%%%%%%%%%%%%%%%%%%%%%%%%%%%%%%%%%%

%%%%%%%%%%%%%%%%%%%%%%%%%%%%%%%%%%%%%%%%%%%%
\begin{figure}
\centering
\hspace{-0.5cm}
\includegraphics[width=8.4cm]{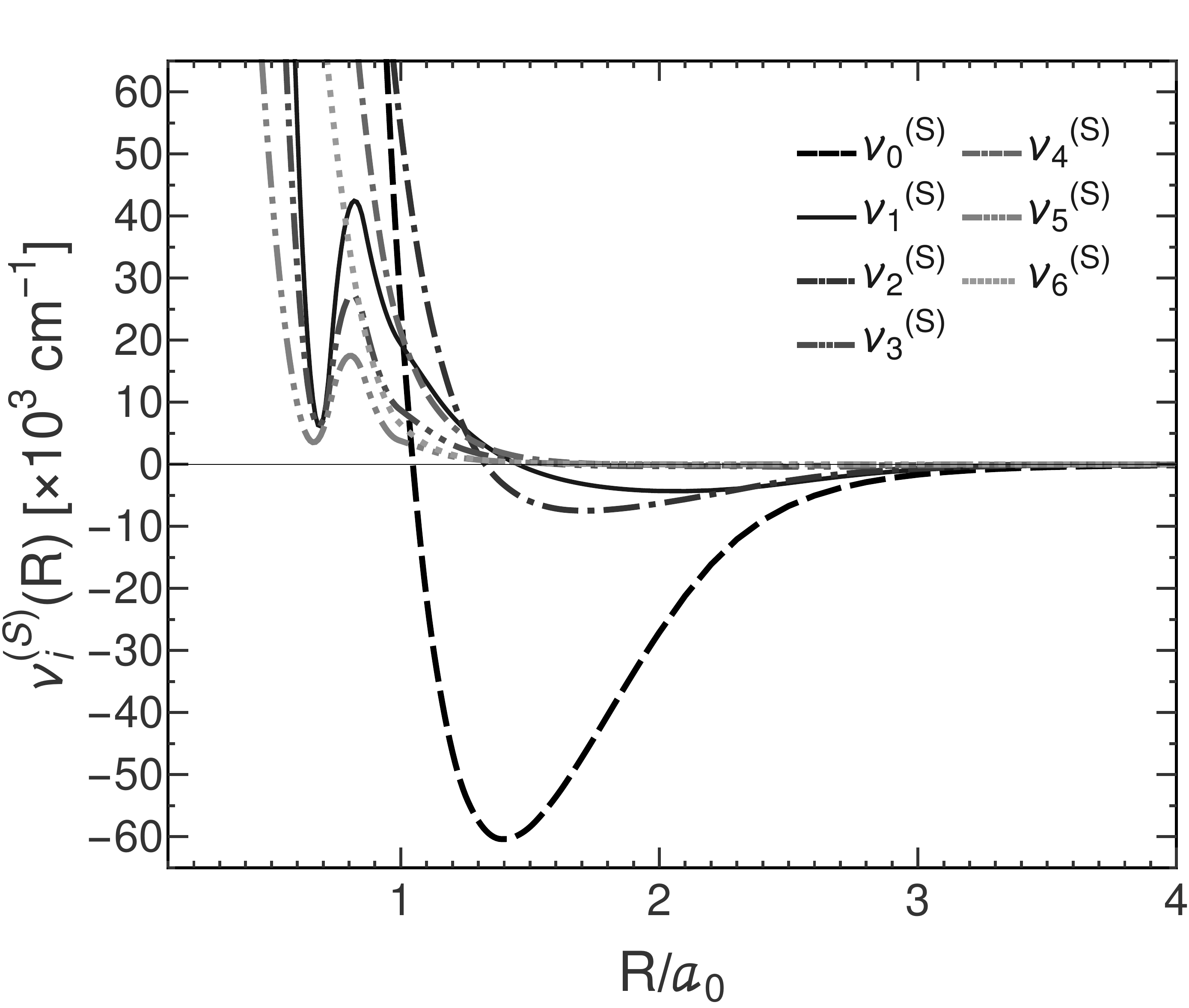} 
\includegraphics[width=8.4cm]{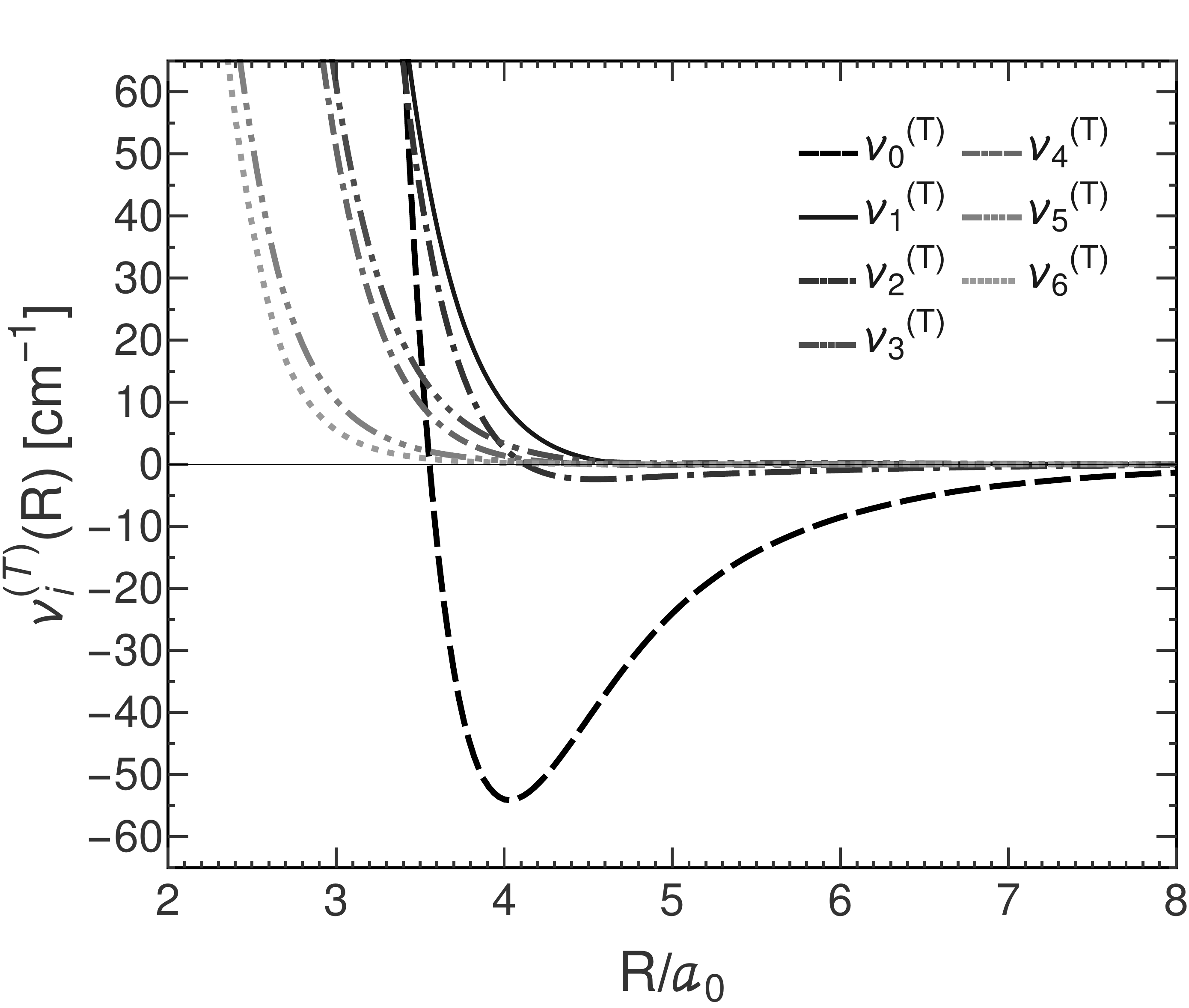}
\hspace{-0.5cm}
\caption{The radial variation of the first seven Legendre coefficients of the singlet (left panel) and triplet (right panel) components of the potential. Note that   the potential well depth  of $V_{\lambda=0}^{(S)}$  is a factor of $\sim$ 10$^3$ greater than the well depth  of $V_{\lambda=0}^{(T)}$.}
\label{fig:PES_S}
\end{figure}
%%%%%%%%%%%%%%%%%%%%%%%%%%%%%%%%%%%%%%%%%%%%

%At a first glance one would think that this does not reflect the vast difference between the depth of the potential well in both parts.

From the above results, one can see that the contributions of the singlet and triplet components of the potential to the depolarization and transfer of polarization cross-sections are not very different from each other despite the vast difference between the  singlet and triplet   interaction potentials. 
To investigate this point, we begin by identifying the radial range to which cross-sections are most sensitive for each of the two potential components. To do so, we add an isotropic local perturbation to the interaction potential $V(R,\theta)$, and thus $V(R,\theta)$ is multiplied by a Gaussian magnification
factor of the form~\footnote{Due to the mild anisotropy present in the potential, one should in principle device a $\theta$-dependent perturbation which can slightly shift the radial range of sensitivty.} 
\begin{eqnarray} \label{eq:sigma_perturbed}
G(R) = 1.+ \exp \left[ -10 (R - R_0)^2 \right]  
\end{eqnarray} 
where $R_0$ refers to the center of the Gaussian perturbation. We vary $R_0$ to scan the entire integration range.
%
%

%
%In fact we will explain this point by showing that, in the region of the potential curve which is decisive for the calculation of the (de)polarization rates,  the triplet and singlet potentials  are not completely different from each other. To investigate this point, we introduce an isotropic local perturbation to the interaction potential $V(r,\theta)$, and thus $V(r,\theta)$ is multiplied by a Gaussian magnification factor of the form 
%\begin{eqnarray} \label{eq:sigma_perturbed}   
%G(r) = 1.+ \exp \left[ -10 (r - r_0)^2 \right]  
%\end{eqnarray} 
%Thus, we can probe the sensitivity of cross-sections to the radial shape of the potential.
We find that the total cross-sections calculated with the perturbed potential components differ significantly from those calculated with the unperturbed potential only when $5 \lesssim R_0 \lesssim 11$ for the singlet component and $2 \lesssim R_0 \lesssim 14$ for the triplet component (see Figure~\ref{fig:sigma_s_perturbed}).  As it can be seen from Figure~\ref{fig:sigma_s_perturbed}, these ranges slightly shift to smaller radial distances as the energy of collision increases.
We note here that for the triplet part of the potential, the radial range where the potential is sensitive to perturbations, is wider compared to that for the singlet component (see Figure~\ref{fig:sigma_s_perturbed}).
By carefully examining the cross-sections calculated with the perturbed and with unperturbed potentials, we find that for the singlet case (the potential component that has a very deep well), the contributions to the cross-sections do not come from the bottom of the well but rather from the part of its far side where the potential starts curving down.
%This is the radial range in which cross-sections are most sensitive to potential perturbations. 
This part of the potential is usually called  ``intermediate region'',  since neither  close-range distances  nor long-range distances are contained in this region.
Interestingly,
in the intermediate region of the singlet part of the potential,
%   where the cross-sections are sensitve to the shape of the potential, 
the difference between the two potential components is not as severe as the difference between the well depth of the two potential components.
%
%this is due to the shallowness of the potential well in the triplet case. In other words, apart from the repulsive wall, values of the singlet and triplet potentials are not vastly different from each other almost everywhere.

Let us expand the potential in the basis of Legendre polynomials,
\begin{eqnarray} \label{V_i_expan1}
V^{(i)} (R,\theta) =
       \sum_{\lambda=0}^{\infty} V_\lambda^{(i)} (R,\theta)
     =\sum_{\lambda=0}^{\infty} \mathcal{V}_{\lambda}^{(i)} (R) P_\lambda (\cos \theta) 
\end{eqnarray}
where
\begin{eqnarray} \label{V_i_expan2}
\mathcal{V}_{\lambda}^{(i)} (R) = \int_{-1}^{1} d( \cos \theta) \, V^{(i)} (R,\theta) P_\lambda (\cos \theta) \, ,
\end{eqnarray}
and separately study the effect of each term on the collision process while ignoring the interference between different potential terms.   In Equations \ref{V_i_expan1} and \ref{V_i_expan2},  $i=S,T$  refers to the singlet and triplet components respectively. Figures~\ref{fig:PES_S} shows the radial dependence of the first seven coefficients of the potential expansion in terms of Legendre functions for the singlet (left panel) and triplet (right panel) parts.
As can be   seen, the depths of the potential wells drastically decrease with increasing $\lambda$.
Further the difference between the Legendre coefficients of the singlet and triplet potential components drastically decreases as $\lambda$ increases.

Separately feeding the different potential terms $V^{(i)}_{\lambda} (R,\theta)$ to MOLSCAT while fixing the energy of collisions to $E=5000$~cm$^{-1}$, we obtain the cross-sections $\sigma(0 \rightarrow L)$,  which we show in Table~\ref{tab:sigma0l}.

\begin{table} 
\centering 
\caption{Scattering cross-sections $\sigma(0 \rightarrow L)$ in \AA$^2$, for collsion energy $E=5000$~cm$^{-1}$.}
\begin{tabular}{ |p{1.05cm}||p{.63cm}|p{.63cm}|p{.63cm}||p{.63cm}||p{.63cm}|p{.63cm}|p{.63cm}||p{.63cm}||p{.63cm}|p{.63cm}|p{.63cm}||p{.63cm}||p{.63cm}|p{.63cm}|} 
\hline
\hline
\!\!\!\! $\sigma(0 \!\! \rightarrow \!\! L)$  & $V_0^{(S)}$ &  $V_0^{(T)}$ & $V_1^{(S)}$ & $V_1^{(T)}$ & $V_2^{(S)}$ & $V_2^{(T)}$ & $V_3^{(S)}$ & $V_3^{(T)}$ & $V_4^{(S)}$ & $V_4^{(T)}$ & $V_5^{(S)}$ & $V_5^{(T)}$ & $V_6^{(S)}$ & $V_6^{(T)}$\\
\hline
$L=0$   & $131.85$ & $71.709$ & $39.806$ & $24.203$ & $31.213$ & $24.013$ & $7.144$ & $13.382$ & $6.153$ &  $13.097$ & $3.906$ & $1.553$ & $4.256$ & $1.110$\\
$L=1$  & $0.000$ & $0.000$ & $6.555$ & $5.306$ & $2.874$ $\times \!  10^{\text{-}5}\!$ &  $0.000$ & $0.301$ & $0.727$ & $0.012$ & $0.000$ & $0.034$ & $0.049$ & $0.000$ & $0.000$\\
$L=2$  & $0.000$ & $0.000$ & $3.592$ & $3.046$ & $6.097$ & $5.203$ & $0.545$ & $0.355$ & $0.310$ & $0.784$ & $0.066$ & $0.149$ & $0.101$ & $0.162$\\
$L=3$  & $0.000$ & $0.000$ & $2.553$ & $2.211$ & $3.346 $ $ \times \! 10^{\text{-}5}\!$ & $0.000$ & $9.184$ &  $5.023$ & $0.007$ & $0.000$ & $0.138$ & $0.203$ & $0.000$ & $0.000$\\
$L=4$  & $0.000$ & $0.000$ & $1.883$ & $1.762$ & $3.161$ & $2.990$ & $0.417$ & $0.383$ & $3.563$ & $4.718$ & $0.031$ & $0.046$ & $0.158$ & $0.177$\\
$L=5$  & $0.000$ & $0.000$ & $1.393$ & $1.474$ & $1.387 $ $ \times \! 10^{\text{-}6}\!$ & $0.000$ & $0.389$ & $0.774$ & $0.005$ & $0.000$ & $1.157$ & $3.518$ & $0.000$ & $0.000$\\
$L=6$  &  $0.000$ & $0.000$ & $1.290$ & $1.269$ & $2.040$ & $2.143$ & $1.521$ & $1.805$ & $0.343$ & $0.724$ & $0.038$ & $0.034$ & $0.906$ & $3.099$\\
$L=7$  & $0.000$ & $0.000$ &   $1.039$ & $1.115$ & $4.411 $ $ \times \! 10^{\text{-}5}\!$ & $0.000$ & $0.174$ & $0.288$ & $0.005$ & $0.000$ & $0.140$ & $0.215$ & $0.000$ & $0.000$\\
$L=8$  & $0.000$ & $0.000$ & $0.867$ & $0.994$ & $1.526$ & $1.681$ & $0.215$ & $1.010$ & $0.378$ & $1.983$ & $0.035$ & $0.049$ & $0.159$ & $0.167$\\
$L=9$  & $0.000$ & $0.000$ & $0.900$ & $0.894$ & $1.451 $ $ \times \! 10^{\text{-}5}\!$ & $0.000$ & $0.208$ & $0.630$ & $0.004$ & $0.0000$ & $0.046$ & $0.077$ & $0.000$ & $0.000$\\
$L=10$  & $0.000$ & $0.000$ & $0.910$ & $0.811$ & $1.236$ & $1.384$ & $0.041$ & $0.162$ & $0.276$ & $0.877$ & $0.392$ & $0.520$ & $0.088$ & $0.090$\\
$L=11$  & $0.000$ & $0.000$ & $0.769$ & $0.740$ & $1.536 $ $ \times \! 10^{\text{-}5}\!$ & $0.000$ & $0.196$ & $0.987$ & $0.007$ & $0.000$ & $0.024$ & $0.042$ & $0.000$ & $0.000$\\
$L=12$  & $0.000$ & $0.000$ & $0.592$ & $0.677$ & $0.860$ & $1.173$ &  $0.083$ & $0.222$ & $0.111$ & $0.869$ & $0.148$ & $0.113$ & $0.289$ &  $0.411$\\
\hline
\hline
\end{tabular}
\label{tab:sigma0l}
\end{table}

\begin{figure}
\centering
%\hspace{-0.7cm}
\includegraphics[width=8.0cm]{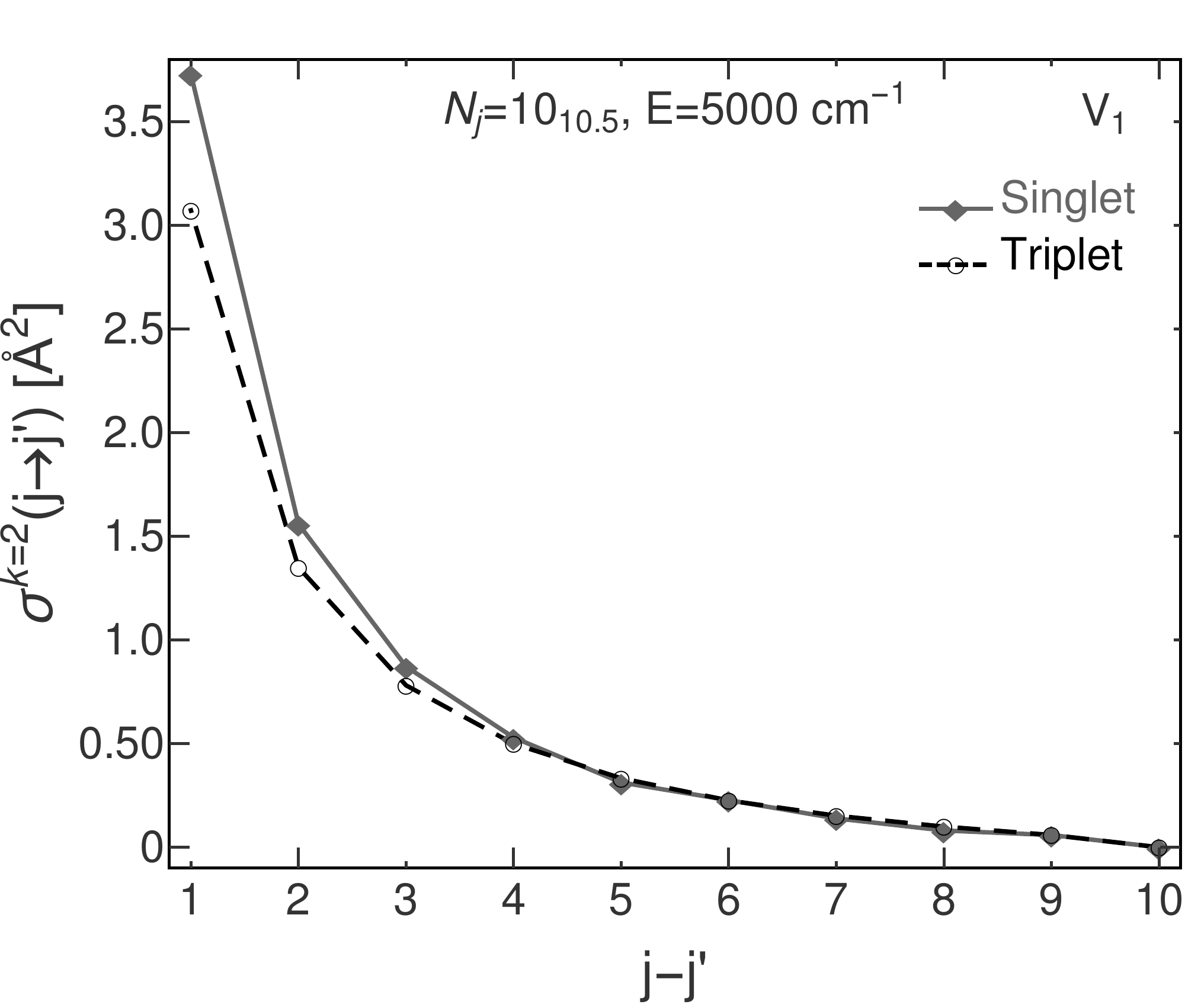} 
\includegraphics[width=8.0cm]{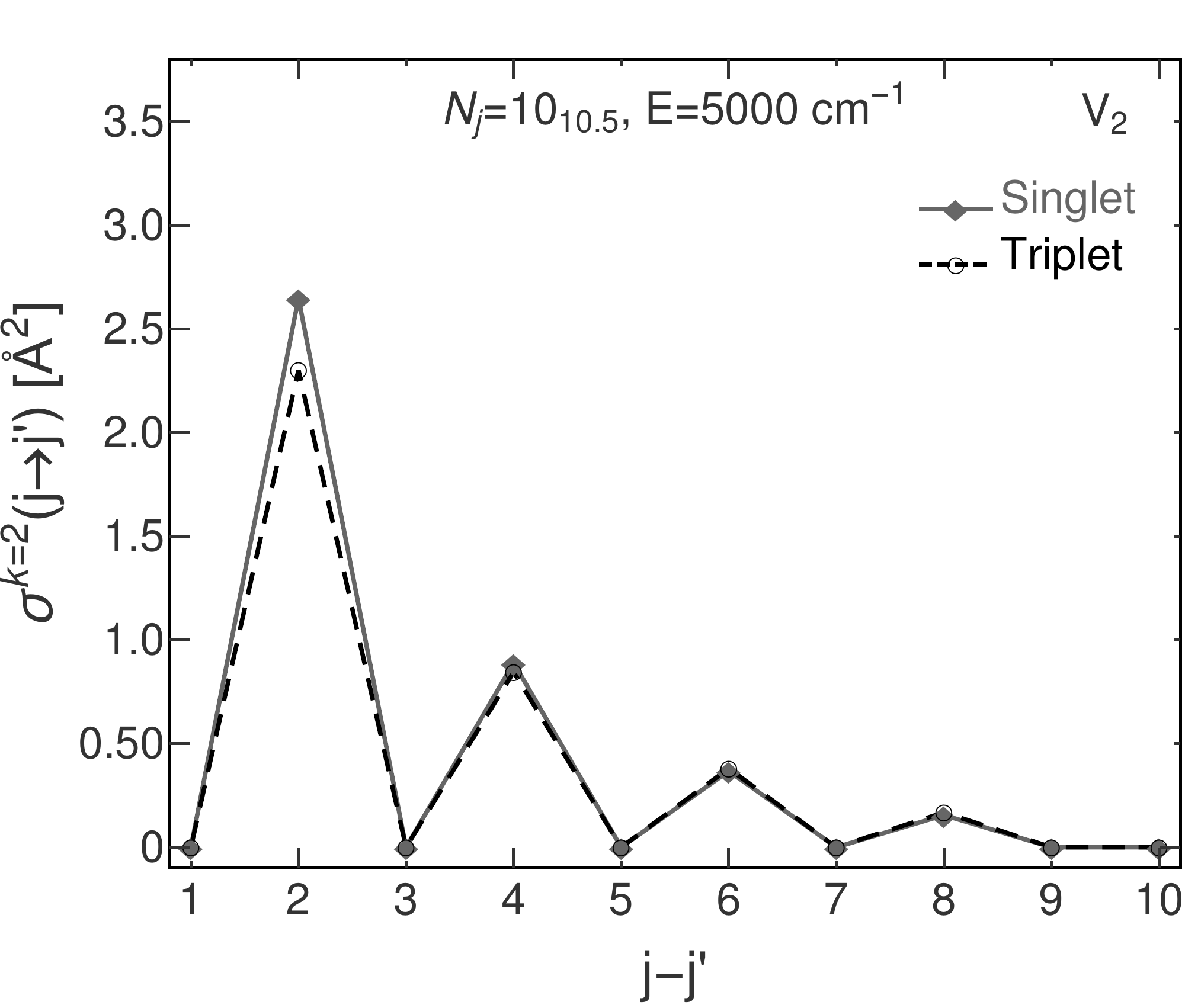}\\
\includegraphics[width=8.0cm]{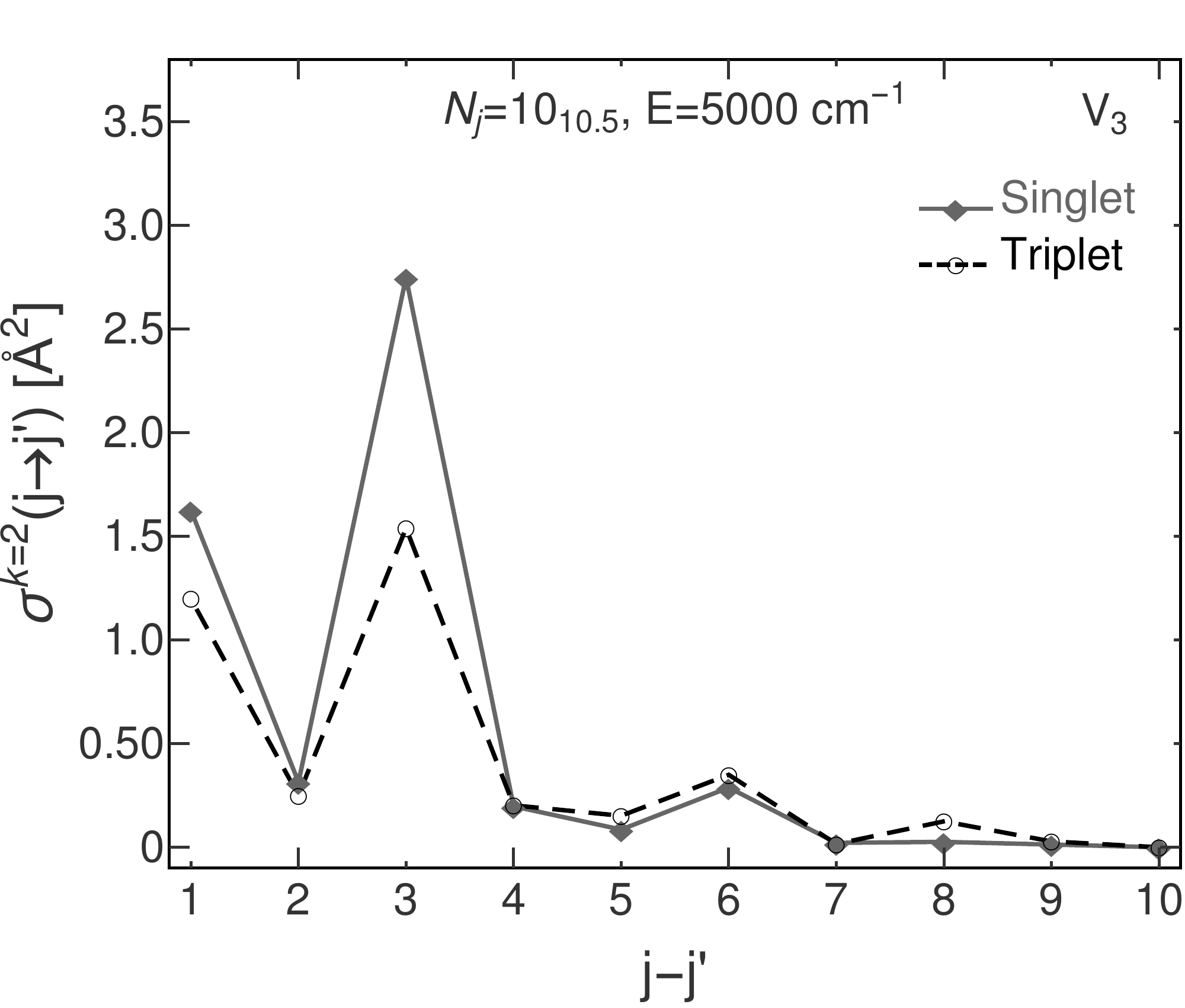}  
\includegraphics[width=8.0cm]{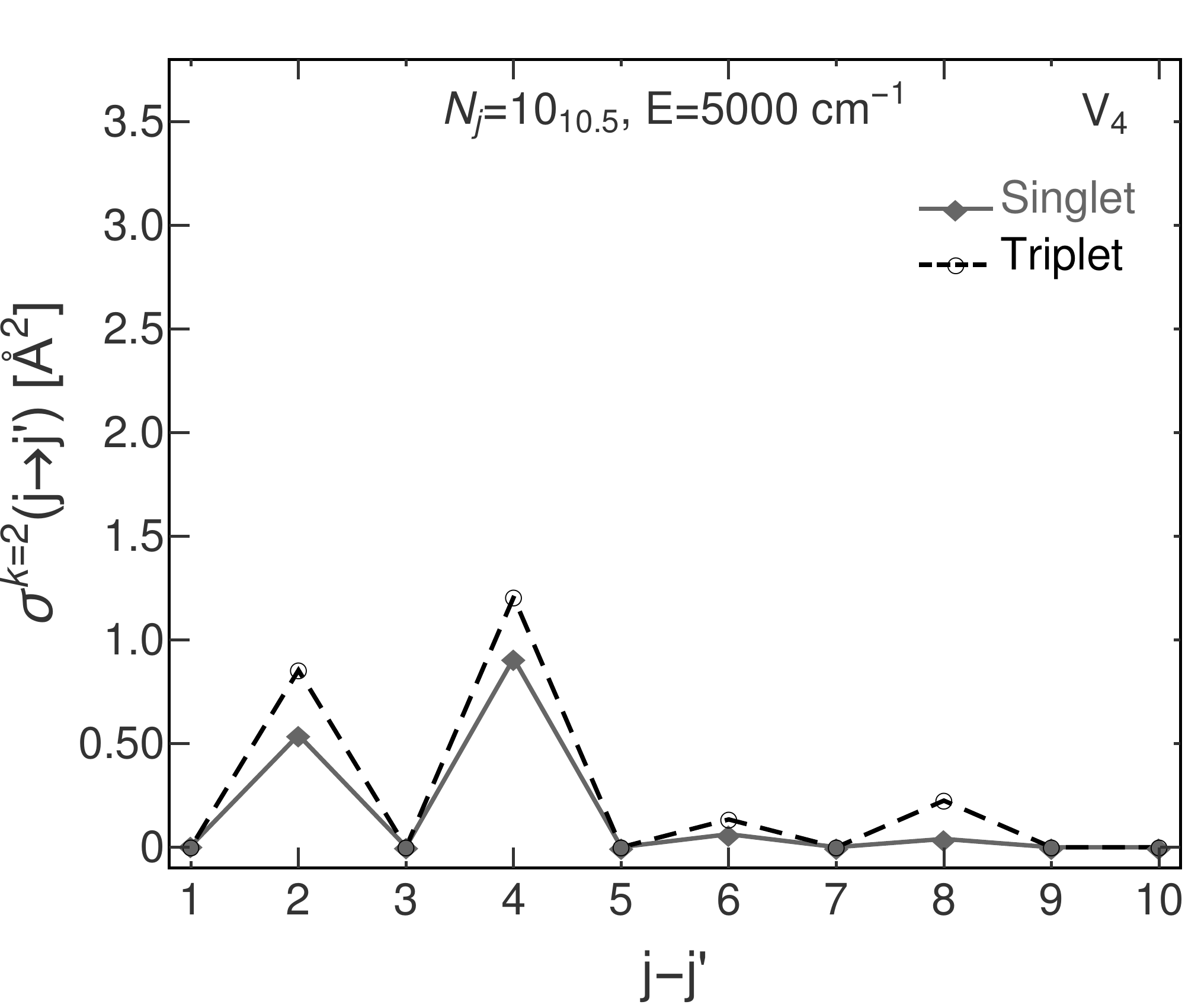} \\
\includegraphics[width=8.0cm]{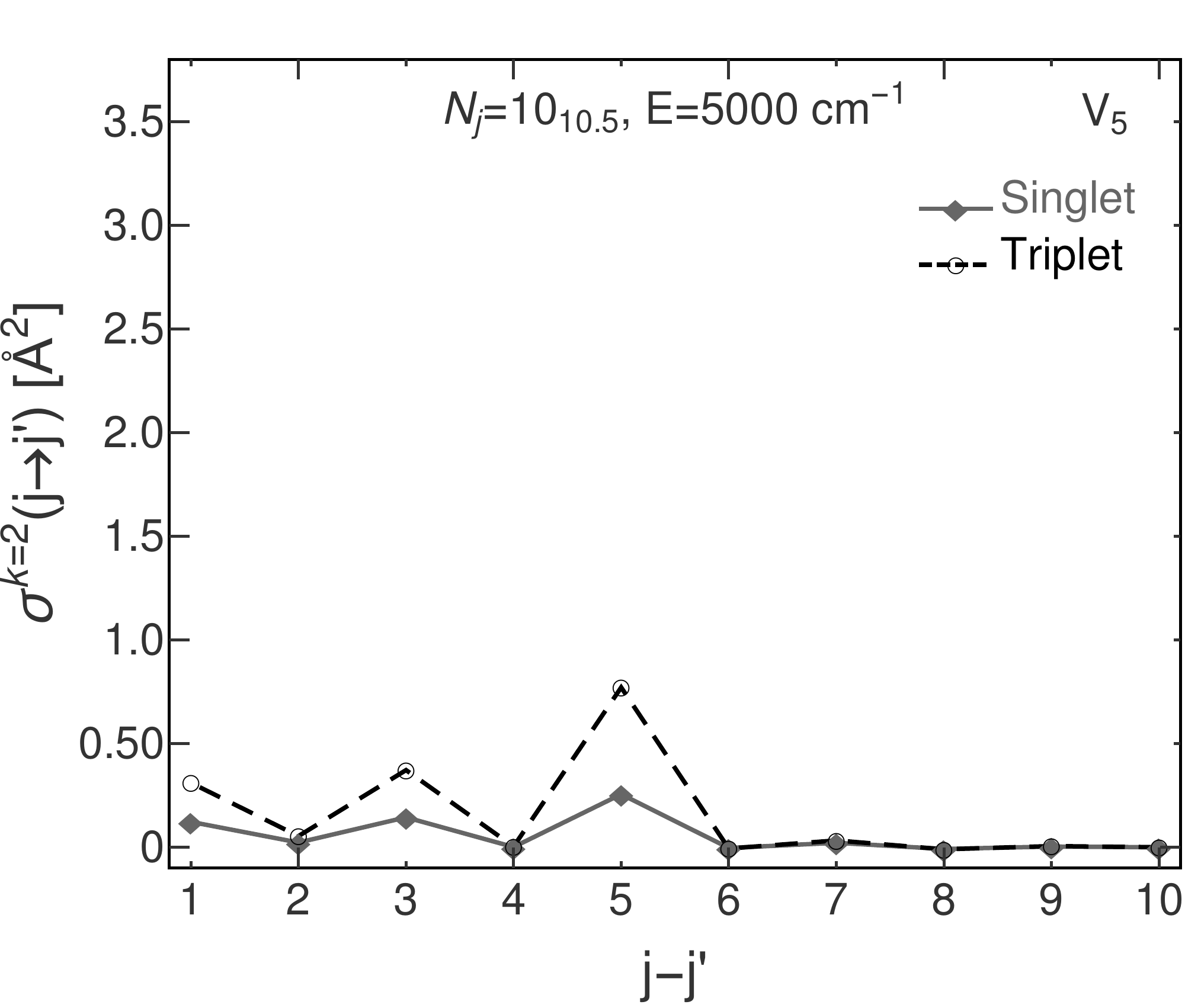}  
\includegraphics[width=8.0cm]{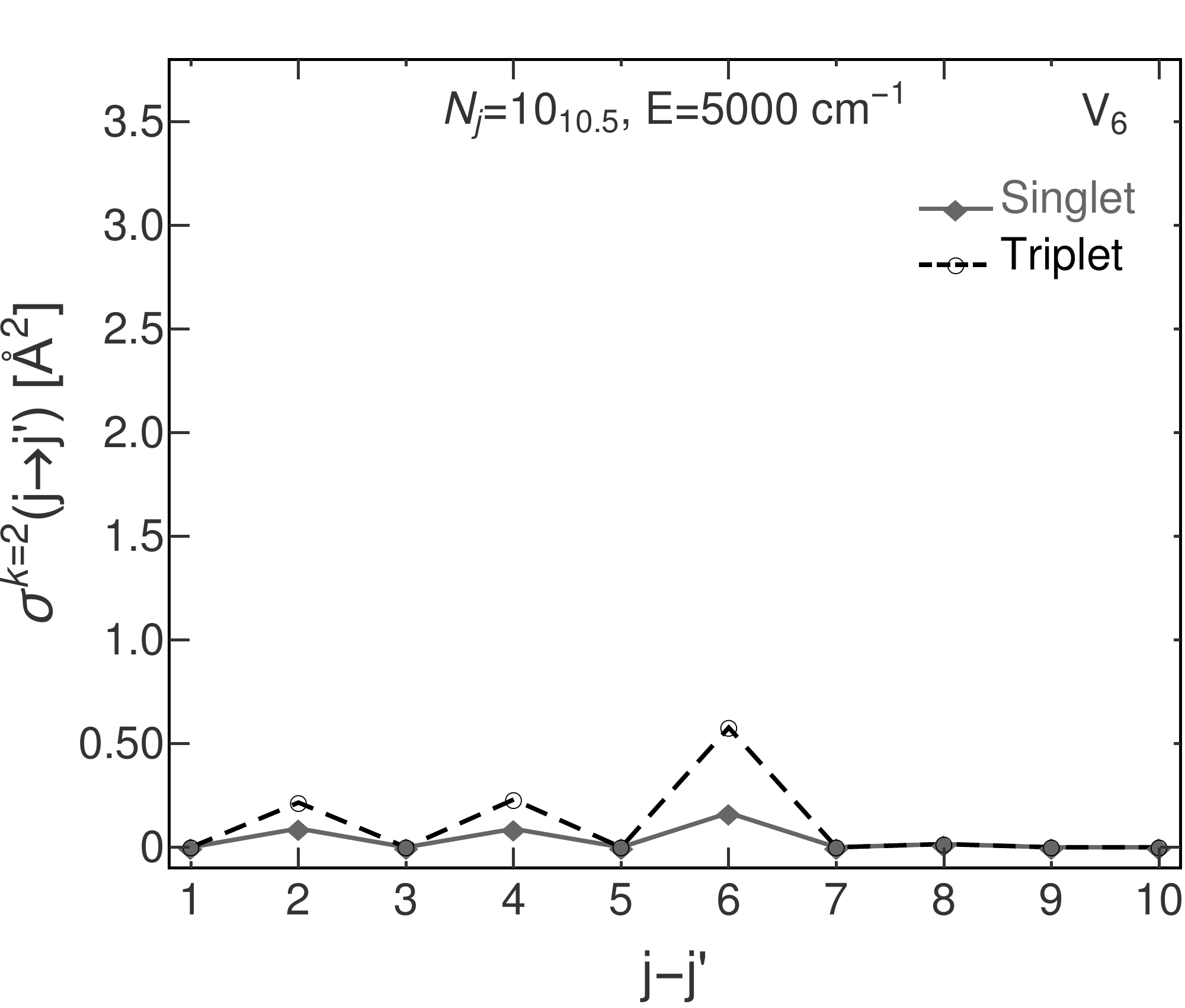}
\caption{Contributions of the different Legendre terms, $V^{(S,T)}_{\lambda}, \, (\lambda=1,2,\cdots \!,6)$, of the singlet (solid gray lines) and triplet (dashed black lines) components of the potential to the transfer of polarization cross-sections for the level $N_j=10_{10.5}$ and $k=2$.}
\label{fig:V1_ST}
\end{figure}

It can be   seen from Table~\ref{tab:sigma0l}
that the components $V_0^{(S,T)}$, which contribute the most to the depth of the potential well, 
contribute only to $\sigma(0 \!\rightarrow\! L \!=\! 0)$
and do not lead to any rotational excitation of the CN molecule
[i.e. they do not contribute to $\sigma(0 \!\rightarrow\! L \!>\! 0)$]
due to their isotropic nature.
It can also be noticed
that for almost all $V_\lambda^{(S,T)}$'s, the largest contribution goes to the $\sigma(0 \!\rightarrow\! L \!=\! 0)$ channel.
In this particular channel,
the singlet contribution is significantly larger than the contribution of the triplet component of the potential for the first few $V^{(i)}_\lambda$ Legendre terms.
For example,
the singlet cross-section is roughly 85\% larger than the triplet cross-section for the $\lambda=0$ term,
which reflects the difference between   $V_0^{(S)}$ and $V_0^{(T)}$. 
Nevertheless,
the difference between the singlet and the triplet contributions to the $\sigma(0 \!\rightarrow\! L \!=\! 0)$ channel decreases as the difference between the two potential components decreases with increasing $\lambda$.
We also note that for the channels with $L>0$, the cross-sections $\sigma(0 \!\rightarrow\! L)$ are almost generally much smaller than the corresponding $\sigma(0 \!\rightarrow\! L \!=\! 0)$.
Moreover,
for a given $V_\lambda^{(S,T)}$ term, the cross-sections are relatively large when $L= m \lambda$ with $m$ being an integer. These cross-sections tend to decrease as $L$ increases.
Furthermore, the difference between the singlet and triplet cross-sections for a given $\lambda$ tends to decease as $L$ increases.

Now since $\sigma(0 \!\rightarrow\! L \!=\! 0)$ does not contribute to depolarization nor transfer of polarization cross-sections,
only the $\sigma(0 \!\rightarrow\! L \!>\! 0)$
arising from the less-different Legendre terms $V_{\lambda>0}^{(S)}$ and $V_{\lambda>0}^{(T)}$ contribute to the depolarization and transfer of polarization cross-sections.
This explains why
the singlet and triplet  depolarization and transfer of polarization rates 
are not very different from each other despite the vast difference
between the singlet and triplet components of the potential.
% well depths  of $V_{\lambda>0}^{(S)}$ and $V_{\lambda>0}^{(T)}$ (Figure \ref{fig:PES_S}). 

Figure~\ref{fig:V1_ST} compares
the contributions to the tranfer of polarization for the level $N_j=10_{10.5}$
from the first six Legendre terms
of the singlet potential terms, $V^{(S)}_{\lambda}, \, (\lambda=1,2,\cdots \!,6)$,
to the corresponding triplet ones, $V^{(T)}_{\lambda}, \, (\lambda=1,2,\cdots \!,6)$,
as functions of $\Delta j=j-j'$
and for the collison energy,
$E=5000$~cm$^{-1}$~\footnote{Here we do not show the contribution of the isotropic Legendre terms $V^{(S,T)}_{0}$ as they are identically zero.}.
It can be concluded
that the singlet and triplet contributions are not very different from each other.

\section{Solar implications}

 Let us now briefly consider the implication of our results for the solar CN molecule. 
To estimate the effect of isotropic collisions,
we compare the collisional depolarization rates $D^2$ of the CN ground state ($X^2\Sigma$),    for typical photospheric Hydrogen density ($n_{H}  =  10^{15}-10^{16}$    cm$^{-3}$), to the inverse lifetime (=$\frac{1}{t_{life}}$) of the lower levels of some representative lines. We show  that the rates $D^2$ dominate the radiative  rates for $n_{H} = 10^{16}$    cm$^{-3}$ and thus all rotational levels  of the lower electronic state $X^2\Sigma$ 
are linearly depolarized.
We notice that $\frac{1}{t_{life}}$=$B_{\ell u} I(\lambda)$,
where
$B_{\ell u} = (g_u/g_{\ell}) (c^2/2h\nu_{u \ell}^3) A_{u \ell}$
denotes the Einstein coefficient for absorption;  $A_{u \ell}$ is the transition probability per unit time for spontaneous emission, $g_u$ and $g_{\ell}$ are the statistical weights of upper and lower levels, $h$ is the Planck's constant and $c$ is the velocity of light. In addition,
\begin{eqnarray}
I (\lambda)= I_{min}(\lambda) I_c(\lambda)
\end{eqnarray}
is the line intensity with $I_{min}(\lambda)$ being the relative intensity of the  line center and $I_c$ being the absolute continuum intensity at disk center.

In Table~\ref{tab:sol_impli}, we show some selected lines of the $B^{2}\Sigma-X^{2}\Sigma$ system of CN along with the corresponding values of radiative excitation rates, $B_{\ell u} I(\lambda)$,  and the linear depolarization rates, $D^{k=2}(j_{\ell})$, calculated at the effective photospheric temperature, $T_{\rm eff} \!=\! 5778$~K, and at typical values of Hydrogen density $n_{H} \!=\! 10^{15} {\rm cm}^{\text{-}3}$ and $n_{H} \!=\! 10^{16} {\rm cm}^{\text{-}3}$ in the photosphere.
The relative intensity of the absorption lines are taken from the solar atlas of Delbouille et al. (1972) whereas the corresponding absolute continuum values are interpolated from the data given in (Allen \& Cox 1999). The values of the Einstein $A$ coefficients, Land\'{e} factors $g_{j_{\ell}}$ and $j_{\ell}$ are taken from  Berdyugina  (2009, private communication).

It is obvious that for $n_{H} \!=\! 10^{16} {\rm cm}^{\text{-}3}$ all the rotational levels of the CN~$X^{2}\Sigma$ state are linearly depolarized since $D^{k=2}(j_{\ell},n_{H} \!=\! 10^{16} {\rm cm}^{\text{-}3}) \gg B_{\ell u} I(\lambda)$. This is also true  in the case $n_{H} \!=\! 10^{15} {\rm cm}^{\text{-}3}$ especially for rather small  $j_{\ell}$; however, for sufficiently large $j_{\ell}$ the radiative excitation and linear depolarization rates of the  CN~$X^{2}\Sigma$ state are comparable. Hence one has to take into account the depolarization rates when solving the statisitcal equilibrium equation for the polarization of observed lines. 
%
%%%%%%%%%%%%%%%%%%%%%%%%%%%%%%%%%%%%%%%%%%%%
\begin{table}
\centering
\caption{\bf Comparison between the linear depolarization rates $D^2$ of the CN~$X^{2}\Sigma$ state to its inverse lifetime  $\frac{1}{t_{life}}$=$B_{\ell u} I(\lambda)$. We also compare the $B_{\ell u} I(\lambda)$ with the values $(\omega_L \vert g_{j_{\ell}} \vert)^{\text{-}1}$ which estimate the Hanle depolarization. }
\begin{tabular}{|c|c|c|c|c|c|c|c|c|c|} 
\hline
\hline
$\lambda \, (\angstrom)$ & $I_{min}$ &  $I_{c} \, (10^{\text{-}5} {\rm erg} \, {\rm cm}^{\text{-}2}$ & $A_{u \ell}$ & $B_{\ell u} I(\lambda)$ & \multicolumn{2}{|c|}{$\omega_{L} \vert g_{j_{\ell}} \vert \, (10^{3} {\rm s}^{\text{-}1})$} & $j_{\ell}$ & 
\multicolumn{2}{|c|}{$D^{k=2}(j) \ ({\rm s}^{\text{-}1})$}
\\
 &  &  ${\rm s}^{\text{-}1}  {\rm sr}^{\text{-}1} {\rm Hz}^{\text{-}1})$ & $(10^{6}{\rm s}^{\text{-}1})$ &  $({\rm s}^{\text{-}1})$ & ${\rm B} \!=\! 10 {\rm G}$ & ${\rm B} \!=\! 100 {\rm G}$ &  & $n_{\rm H} \!=\! 10^{15} {\rm cm}^{\text{-}3}$  &  $n_{\rm H} \!=\! 10^{16} {\rm cm}^{\text{-}3}$
\\
\hline
$3839.136$ & $0.1920$ & $1.0462$ & $7.0401$ & $2014.16$ & $2286.5$ & $22865$ & $37.5$ & $7864.13$ & $78641.3$
\\
$3850.178$ & $0.1707$ & $1.0479$ & $6.9783$ & $1793.18$ & $2981.2$ & $29812$ & $27.5$ & $12359.3$ & $123593$
\\
$3862.692$ & $0.6564$ & $1.0497$ & $7.0958$ & $7092.58$ & $1231.2$ & $12312$ & $72.5$ & $2446.76$ & $24467.6$
\\
$3870.871$ & $0.2097$ & $1.0510$ & $7.0811$ & $2278.23$ & $1407.1$ & $14071$ & $62.5$ & $3228.44$ & $32284.4$
\\
$3871.372$ & $0.1425$ & $1.0510$ & $6.3860$ & $1396.82$ & $13502.9$ & $135029$ & $4.5$ & $79432.9$ & $794329$
\\
$3879.707$ & $0.3365$ & $1.0523$ & $7.0592$ & $3674.10$ & $1890.7$ & $18907$ & $46.5$ & $5491.44$ & $54914.4$
\\
$3880.681$ & $0.2635$ & $1.0524$ & $7.2926$ & $2974.96$ & $7650.9$ & $76509$ & $11.5$ & $32029.9$ & $320299$
\\
$3880.784$ & $0.2763$ & $1.0524$ & $7.0763$ & $3027.07$ & $2022.6$ & $20226$ & $43.5$ & $6156.94$ & $61569.4$
\\
$3883.114$ & $0.1568$ & $1.0528$ & $7.1339$ & $1675.55$ & $3904.6$ & $39046$ & $22.5$ & $15886.1$ & $158861$
\\
\hline
\hline
\end{tabular}
\label{tab:sol_impli}
\end{table}
%%%%%%%%%%%%%%%%%%%%%%%%%%%%%%%%%%%%%%%%%%%%
%

  We also consider the effect of the photospheric turbulent magnetic field on the polarization of the CN ground state,  $X^{2}\Sigma$. It is well known that the polarization is sensitive to Hanle depolarization   only if the magnetic field value is around a critical value B$_c$ (more precisely the magnetic field has a value in the window $\sim 0.1 {\rm B}_c - 10 {B}_c$). In other words, Hanle effect is relevant if the time-life of the level under consideration [$ \sim (B_{\ell u} I)^{\text{-}1}$ for the ground state] is of order $(\omega_L \vert g_{j} \vert)^{\text{-}1}$ where $\omega_L \!=\! 8.79 \!\times\! 10^{6}$B denotes Larmor angular frequency with B being the magnetic field stength in Gauss.

In Table~\ref{tab:sol_impli}, we show the values of $\omega_L \vert g_{j_{\ell}} \vert$ calculated at ${\rm B} \!=\! 10 {\rm G}$ and ${\rm B} \!=\! 100 {\rm G}$. It is clear  that $\omega_L \vert g_{j_{\ell}} \vert \gg B_{\ell u} I$ in all cases. Therefore, for typical values of the turbulent magnetic field $\sim 10-100$~G, Hanle effect is not efficient for the CN ground sate, CN~$X^{2}\Sigma$.
CN~$X^2\Sigma$  is sensitive to the Hanle effect of very weak magnetic field strength since it is a long lived level and thus the saturation regime of the Hanle effect on its linear polarization is quickly attempted. 
%The the depolarization effect can occurs mainly due to collisions.

\section{Conclusion}
The so-called  second solar spectrum (SSS) of the CN molecule is the observational signature of the  polarization of the CN states (see, e.g.,   Trujillo Bueno et al. 2004, Landi Degl'Innocenti \&  Landolfi 2004). 
Molecular lines arising from transitions between rotational levels can be depolarized by collisions but also by the Hanle effect due to the presence of solar magnetic fields. Therefore, information about Hanle and collisional effects are mixed in the same observable (the polarization state), which makes the interpretation of the observed polarization in terms of magnetic fields very complicated in the absence of collisional data.  

In this paper we provide   depolarization and  polarization transfer rates of the $X \; ^2\Sigma$ state of the CN due to collisions with neutral hydrogen in its ground state  $^2S$. These rates  would be   useful to interpret  CN violet lines
in the second solar spectrum in terms  of solar magnetic field   (Shapiro et al.  2011).   
A detailed discussion of the results is presented we obtain  useful variation  laws of the polarization transfer rates    
with the temperature and the angular momentum $j$.  Solar implications of our results are discussed.

%and 

\section*{Aknowledgements} 
M.D. wishes to thank    the   {\it Universit\'e Le Havre Normandie} for their kind invitation and hospitality. 
 Results of \textit{ab initio} calculations  have been obtained under support of the RSF grant No. 17-12-01395. Y.K. also acknowledges the partial support from RFBR grant No. 18-05-00119. 

%%%%%%%%%%%%%%%%%%%%%%%%%%%%%%%%%%%%%%%%%%%%%%%%%%



%\bibliographystyle{mnras}


%%%%%%%%%%%%%%%%%%%%%%%%%%%%%%%%%%%%%%%%%%%%%%%%%%

%%%%%%%%%%%%%%%%% APPENDICES %%%%%%%%%%%%%%%%%%%%%

%%%%%%%%%%%%%%%%%%%%%%%%%%%%%%%%%%%%%%%%%%%%%%%%%%

% Don't change these lines
%\bsp	% typesetting comment
\label{lastpage}
\end{document}